\documentclass[11pt]{article}
\usepackage{jheppub}
\usepackage{dsfont}
\usepackage{amsmath,amssymb,amscd,amsfonts,mathtools}

\usepackage[toc,page]{appendix}
\usepackage{epsfig}
\usepackage{epstopdf}
\usepackage{latexsym}
\usepackage{graphicx}
\usepackage{subfig}
\usepackage{booktabs}
\usepackage{bbm}
\usepackage{color}
\usepackage{physics}
\usepackage{tensor}
\usepackage{tikz}
\usepackage{tcolorbox} % Lets me put 'observations' in nice colored boxes. 
\usetikzlibrary{matrix}
\usetikzlibrary{decorations.markings,calc,shapes,decorations.pathmorphing,patterns}
\usetikzlibrary{positioning}
\makeatletter
\def\@fpheader{\relax}
\makeatother

\definecolor{markgreen}{RGB}{230,243,230}

\subheader{\begin{flushright}
\end{flushright}}

\title{Entanglement Phase Structure of a Holographic BCFT in a Black Hole Background}

\author{Hao Geng$^{a,b}$, 
Andreas Karch$^{b,c}$, Carlos Perez-Pardavila$^{c}$, Suvrat Raju$^d$, Lisa Randall$^a$, Marcos Riojas$^{c}$, Sanjit Shashi$^{c}$}
\affiliation{$^a$Harvard University, 17 Oxford St., Cambridge, MA, 02139, USA.}
\affiliation{$^b$Department of Physics, University of Washington, Seattle, WA, 98195-1560, USA.}
\affiliation{$^c$Theory Group, Department of Physics, University of Texas, Austin, TX 78712, USA.}
\affiliation{$^d$International Centre for Theoretical Sciences, Tata Institute of Fundamental Research, Shivakote, Bengaluru 560089, India.}

\emailAdd{gengphysics666@gmail.com, 
karcha@utexas.edu, cjp3247@utexas.edu}
\emailAdd{suvrat@icts.res.in, randall@g.harvard.edu}
\emailAdd{marcos.riojas@utexas.edu, sshashi@utexas.edu}

\abstract{We compute holographic entanglement entropy for subregions of a BCFT thermal state living on a nongravitating black hole background. The system we consider is doubly holographic and dual to an eternal black string with an embedded Karch-Randall brane that is parameterized by its angle. Entanglement islands are conventionally expected to emerge at late times to preserve unitarity at finite temperature, but recent calculations at zero temperature have shown such islands do not exist when the brane lies below a critical angle. When working at finite temperature in the context of a black string, we find that islands exist even when the brane lies below the critical angle. We note that although these islands exist when they are needed to preserve unitarity, they are restricted to a finite connected region on the brane which we call the atoll. Depending on two parameters---the size of the subregion and the brane angle---the entanglement entropy either remains constant in time or follows a Page curve. We discuss this rich phase structure in the context of bulk reconstruction.}

\begin{document}	
\maketitle
\flushbottom

%%%%%%%%%%%%%%%%%
\section{Introduction}
%%%%%%%%%%%%%%%%%

The \textit{Ryu-Takayanagi (RT) prescription} \cite{Ryu:2006bv,Ryu:2006ef,Hubeny:2007xt}, and its generalization to incorporate quantum effects \cite{Faulkner:2013ana,Lewkowycz:2013nqa,Engelhardt:2014gca}, have proven to be powerful tools in the AdS/CFT correspondence \cite{Maldacena:1997re,Gubser:1998bc,Witten:1998qj}. These techniques yield remarkably simple formulas for the entropy of subregions of strongly coupled conformal field theories with bulk gravitating AdS duals, and they have since been extended to compute entropy in conformal field theories on spaces with boundaries (BCFTs). 

Some strongly coupled BCFTs admit multiple dual descriptions and are called ``doubly holographic" \cite{Karch:2000gx,Takayanagi:2011zk,Fujita:2011fp}. The dual descriptions comprise AdS$_{d+1}$ gravity containing a subcritical end-of-the-world \textit{Karch-Randall (KR) brane} \cite{Randall:1999vf,Karch:2000ct} and an ``intermediate" description where brane-localized AdS$_d$ gravity coupled to a CFT is connected to a nongravitating bath with the same CFT \cite{Almheiri:2019psf,Penington:2019npb,Almheiri:2019hni,Almheiri:2019psy,Almheiri:2019yqk}. The $(d-1)$-dimensional boundary of the BCFT is sometimes called a \textit{defect}.

In this paper, we will study a black-string state in the bulk theory illustrated in Figure \ref{fig:holography}. This is dual to an AdS$_d$ black hole coupled to a bath but it is also described by a CFT in a thermal state that itself resides on a nongravitating eternal black hole background. The braneworld black-hole horizon is where the black-string horizon connects to the brane and the horizon also extends to the boundary theory. 

Since the BCFT description of this system is entirely nongravitational we expect that, in a generic state, the entanglement entropy of subregions in the dual BCFT will be governed by a universal expression depending on the size of the subregion and its complement \cite{Page:1993df,lubkin1978entropy}. However, the original state of the BCFT at $t = 0$ is not generic; this is reflected in correlations between operators in the left and right wedges of the background black hole. The subregions that we study have support in both of these wedges, and so their initial entropy is sometimes lower than the generic value. Time evolution leads to the loss of correlations between the two sides, and this causes the entanglement entropy of these subregions to approach that of a typical state, leading to a ``Page curve" for the evolution of this entropy with time.

From the $(d+1)$-dimensional bulk perspective, the Page curve is expected to emerge classically because, while the early time entropy is controlled by an extremal \emph{Hartman-Maldacena (HM) surface} \cite{Hartman:2013qma} which crosses the black-string horizon and traverses the growing Einstein-Rosen bridge, the late-time entropy should be controlled by an extremal \emph{island surface} which connects the brane to the conformal boundary and is constant in time \cite{Almheiri:2019yqk}. Such island surfaces are standard RT surfaces ending on the KR brane, but in the $d$-dimensional braneworld these are dual to \textit{quantum extremal surfaces (QES)} \cite{Engelhardt:2014gca,Almheiri:2019hni} which are disconnected from the AdS$_d$ boundary and enclose (semiclassical) ``islands" on the brane.\footnote{Entanglement islands have been studied extensively in the literature \cite{Ling:2020laa,KumarBasak:2020ams, Emparan:2020znc,Caceres:2020jcn,Caceres:2021fuw,Deng:2020ent, Krishnan:2020fer,Balasubramanian:2020coy,Balasubramanian:2020xqf,Manu:2020tty,Karlsson:2021vlh,Wang:2021woy,Miao:2021ual,Bachas:2021fqo,May:2021zyu,Kawabata:2021hac,Bhattacharya:2021jrn,Anderson:2021vof,Miyata:2021ncm,Kim:2021gzd,Hollowood:2021nlo,Wang:2021mqq,Aalsma:2021bit,Ghosh:2021axl,Neuenfeld:2021wbl,Geng:2021iyq,Balasubramanian:2021wgd,Uhlemann:2021nhu,Neuenfeld:2021bsb,Kawabata:2021vyo,Chu:2021gdb,Kruthoff:2021vgv,Akal:2021foz,KumarBasak:2021rrx,Lu:2021gmv,Omiya:2021olc,Ahn:2021chg,Balasubramanian:2021xcm,Li:2021dmf,Kames-King:2021etp,Sun:2021dfl,Hollowood:2021wkw,Miyaji:2021lcq,Bhattacharya:2021dnd,Goswami:2021ksw,Chu:2021mvq,Arefeva:2021kfx,Shaghoulian:2021cef,Garcia-Garcia:2021squ,Buoninfante:2021ijy,Yu:2021cgi,Nam:2021bml,He:2021mst,Langhoff:2021uct,Ageev:2021ipd,Pedraza:2021cvx,Iizuka:2021tut, Miyata:2021qsm,Gaberdiel:2021kkp,Uhlemann:2021itz,Collier:2021ngi,Hollowood:2021lsw,emparan2021holographic,omidi2021entropy,bhattacharya2021bath}, and useful reviews are \cite{Almheiri:2020cfm,Raju:2020smc,Raju:2021lwh,Liu:2020rrn,Nomura:2020ewg,Kibe:2021gtw}.}

The story clearly breaks down if the bulk island surfaces do not exist in the first place,\footnote{Concretely, ``nonexistence" of island surfaces means that the area functional of a codimension-2 bulk surface ending on the brane does not have any extrema. This is qualitatively different from the island surfaces existing but being subleading in the entanglement entropy.} and recent computations have shown \cite{Chen:2020hmv,Geng:2020fxl} that this may happen in the empty AdS$_{d+1}$ bulk for $d > 2$. More precisely, there is a \textit{critical angle} for the brane that depends on $d$, and island solutions do not exist when the brane lies below that angle. This is not necessarily a problem for empty AdS$_{d+1}$ because there will be no Einstein-Rosen bridge, but island solutions are needed to obtain the Page curve when there is a black hole in the bulk. This concern is not unique to the black string model. The critical angle plays an important role in every braneworld model which is asymptotically AdS \cite{Geng:2020fxl}, and there seems to be no guarantee that island solutions exist in all configurations, even when they are needed to preserve unitarity.

To address this issue, this paper considers a tractable geometric configuration in which islands and their properties can be studied numerically. First a KR brane is embedded in an eternal AdS$_{d+1}$ planar black string. The symmetries of the background allow the brane to be inserted at an arbitrary angle using the Israel junction conditions \cite{Israel:1966rt} that uniquely associate the brane tension with a given brane angle. Note that the planar black string is stable in that it does not have the usual Gregory-Laflamme instability \cite{Gregory:1993vy}. Furthermore the horizon extends all the way to the conformal boundary and this is why the dual BCFT state itself resides on a nongravitating black hole background \cite{Marolf:2013ioa}. 

Although the bulk computation of the entropy is simpler than an ab initio computation in the strongly coupled BCFT \cite{Rozali:2019day,Sully:2020pza}, it still often requires the numerical solution of partial differential equations \cite{Almheiri:2019psy,Uhlemann:2021nhu}. In this paper we show how the entropy of subregions for this BCFT state can be computed in the bulk using simple numerical techniques. Our main result is a presentation of the rich phase structure for this system, in which island solutions exist for all values of brane angle and finite subregion size, even below the critical angle. Instead of failing to exist, the semiclassical islands for fixed $\theta_b$ are restricted to a region of the brane which we call the \textit{atoll}, which we illustrate in Figure \ref{fig:RT_cartoon}.

To determine which surface is dominant (i.e. minimizes the area functional) on the initial time slice, one simply computes the area difference between the HM and island surfaces for the same parameters.\footnote{The areas are technically UV-divergent, but the area difference is manifestly UV-finite since all extremal surfaces anchored to the conformal boundary exhibit the same divergence structure \cite{Graham:1999pm}.} We find that subregion entropy depends on both subregion size and brane angle. Depending on the values of these parameters, the initial dominant bulk surface is either an HM surface or the union of an island surface and its thermofield partner.

We obtained some simple rules for the phase structure. One can tune the size of the island by varying either radiation region size or the angle of the brane. Shrinking the radiation region size or decreasing the brane angle decreases the island size (except for the special case when the radiation region is the entire bath and the island covers the whole brane). In Figure \ref{fig:russiandoll} we illustrate the Russian doll rule, which states that nested surfaces can
be constructed by either decreasing the brane angle $\theta_b$ or increasing the anchor point $\Gamma$. We show how this ``Russian doll rule" can be used to qualitatively compare the area difference for different parameter choices. This rule shows that the Page time diverges as the parameters are varied to probe regions near the black-string horizon in the bulk. We comment on this rule in the context of entanglement wedge reconstruction. 

Another interesting feature of this model is the ``constant entropy belt," which is a connected interval on the boundary defined as the values of $\Gamma$ for which island surfaces dominate on the initial time slice. When the angle of the brane $\theta_b$ is increased from $0$ to $\pi$, the belt first appears slightly below the critical angle (at what we call the \emph{Page angle}) and grows monotonically until it covers the entire boundary as $\theta_b \rightarrow \pi$. We find an interesting formula describing this region, which we elaborate on in Section \ref{sec:symmetry}. 
 
\paragraph{Relationship with previous work.}
Before we proceed to our analysis, we pause to clarify how the results of this paper should be interpreted in light of our previous results on gravitating baths and the relationship between massive gravity and islands \cite{Geng:2020fxl, Geng:2021hlu}.

In \cite{Geng:2020fxl}, we showed that the physics of an AdS$_d$ black hole coupled to a bath is qualitatively different when the bath is \textit{gravitating} as opposed to nongravitating. Specifically, for a gravitating bath the Page curve disappears, even if gravity in the bath is parameterically weak (i.e. the ratio of $G_N$ between the ``bath brane" and the ``system brane" is small). This is consistent with the finding that the full fine-grained entropy at ${\cal I}^{+}$ does not follow a Page curve for black holes in asymptotically flat space \cite{Laddha:2020kvp,Raju:2020smc}. The nongravitating bath also induces a mass for the graviton in the region where gravity is dynamical \cite{Aharony:2006hz} as a result of a bulk one-loop effect \cite{Porrati:2003sa}. In \cite{Geng:2021hlu} we showed that this mass causes a breakdown of the Gauss law that is essential for islands to constitute consistent entanglement wedges; in theories of gravity where the Gauss law applies, islands are inconsistent even in perturbation theory.

So we caution the reader that, in contrast to some proposals in the literature \cite{Krishnan:2020oun,Almheiri:2020cfm}, we do not necessarily view the nongravitating bath that appears in this paper as a model for the region far from a black hole in asymptotically flat space. Moreover, the graviton that is localized on the brane is massive. Therefore the Page curve that we discuss below should not be thought of as the Page curve for Hawking radiation in a standard theory of gravity. Our results have a clean interpretation in terms of the entanglement entropy of subregions in the nongravitating BCFT---which is a problem of independent interest and one for which the Page curve is relevant---and we use the bulk theory of gravity simply as a tool to perform this computation.

\paragraph{Overview of the Paper.}

We first discuss the geometrical aspects of our model in Section \ref{seckrreview}, starting with a review of double holography and then delving into the types of extremal surfaces that appear in the black string with a KR brane present. We then discuss the numerics in Section \ref{secardiff}, presenting our results for the phase structure (summarized above) in detail. We conclude with some discussion of our results and potential future directions in Section \ref{secdiscuss}.

%%%%%%%%%%%%%%%%%
\section{The Doubly Holographic Black-String Model \label{seckrreview}}
%%%%%%%%%%%%%%%%%

\begin{figure}
 \centering
 \includegraphics[width=\linewidth]{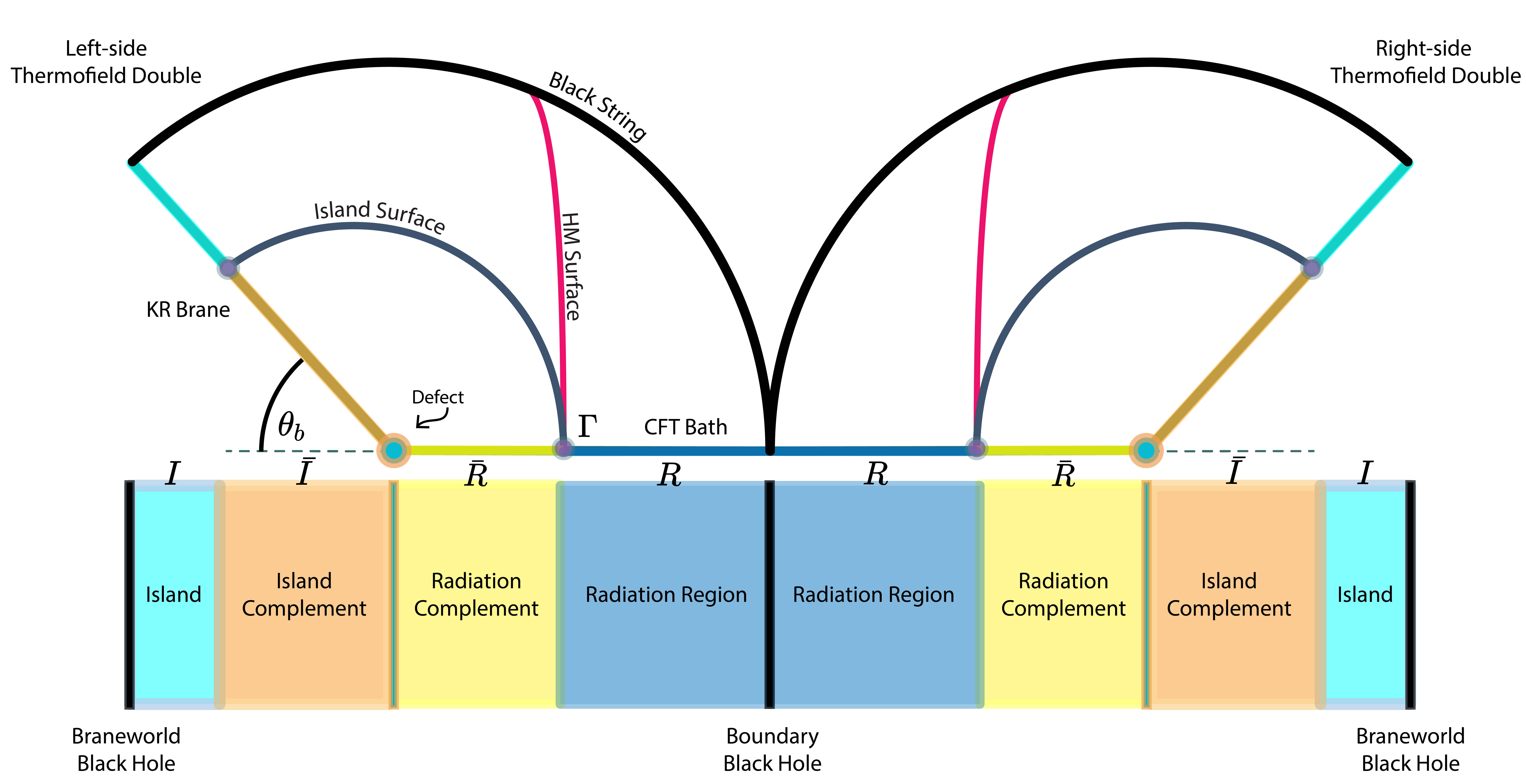}
 \caption{This cartoon illustrates the doubly holographic black-string model, where the region behind the KR brane has been excised. There are three perspectives, with the $\text{AdS}_{d+1}$ gravitating bulk illustrated at the top of the diagram. The $d$-dimensional BCFT and $d$-dimensional $\text{AdS}_d$ brane are depicted in colors which correspond with the intermediate picture indicated at the bottom of the diagram. Notice the island region envelops the black-hole horizon on the brane, and the defect cannot be reached from the island in the intermediate picture without crossing through its complement. The two parameters in the theory are the anchor point $\Gamma$ and brane angle $\theta_b$. Note that ``left/right" has a different meaning than in our previous paper \cite{Geng:2020fxl}.}
\label{fig:holography}
\end{figure}

This section describes general features of the doubly holographic black-string model, illustrated in Figure \ref{fig:holography}. The first two subsections are mostly qualitative and we refer the reader to \cite{Geng:2020fxl} for additional technical details. Readers who are already familiar with the literature may want to skip the overview of doubly holographic models in Section \ref{sec2.1}. The main point is to review the doubly holographic setup and point out that transparent boundary conditions make the graviton on the KR brane massive. 

Section \ref{sec2.2} provides an overview of the doubly holographic black-string model, illustrated in Figure \ref{fig:holography}, and determines the area functional for the codimension-2 surfaces. Section \ref{sec2.3} explains how boundary conditions distinguish the extremal surfaces into two distinct classes. These are the \textit{Hartman-Maldacena (HM) surfaces} \cite{Hartman:2013qma}, which avoid the branes and traverse the Einstein-Rosen bridge, and \textit{island surfaces}, which connect the boundary to the brane and avoid the Einstein-Rosen bridge. Due to the growth of the Einstein-Rosen bridge, the former grow with boundary time and the latter do not. The interplay between these two classes of surfaces ensures the growth of the entanglement entropy will be consistent with unitarity, provided the island surfaces exist. In Section \ref{sec2.4} we review recent work performed independently by our group \cite{Geng:2020fxl} and others \cite{Chen:2020hmv} that revealed that island solutions sometimes \emph{fail to exist} when the boundary has more than two spatial dimensions. This potentially leads to a serious puzzle in any braneworld model.

\subsection{Review of Doubly Holographic KR Models} \label{sec2.1}

While KR models have recently become more visible, they have been well understood for some time as having three equivalent descriptions \cite{Karch:2000ct,Karch:2000gx,Almheiri:2019hni,Takayanagi:2011zk,Fujita:2011fp}:
\begin{itemize}
\item \textbf{Boundary:} $d$-dimensional CFT with a $(d-1)$-dimensional defect ($\text{BCFT}_d$ \cite{Cardy:2004hm,McAvity:1995zd}).
\item \textbf{Bulk:} Einstein gravity in an asymptotically $\text{AdS}_{d+1}$ spacetime containing an $\text{AdS}_d$ Karch-Randall (KR) brane \cite{Karch:2000ct}.
\item \textbf{Intermediate:} $d$-dimensional CFT coupled to ``localized" gravity on the $\text{AdS}_d$ brane \cite{Karch:2000ct}, with transparent boundary conditions between the brane at infinity and a nongravitating $\text{CFT}_d$ (the \textit{bath}) on half space.
\end{itemize}

\begin{figure}
 \centering
 \includegraphics[width=.3\linewidth]{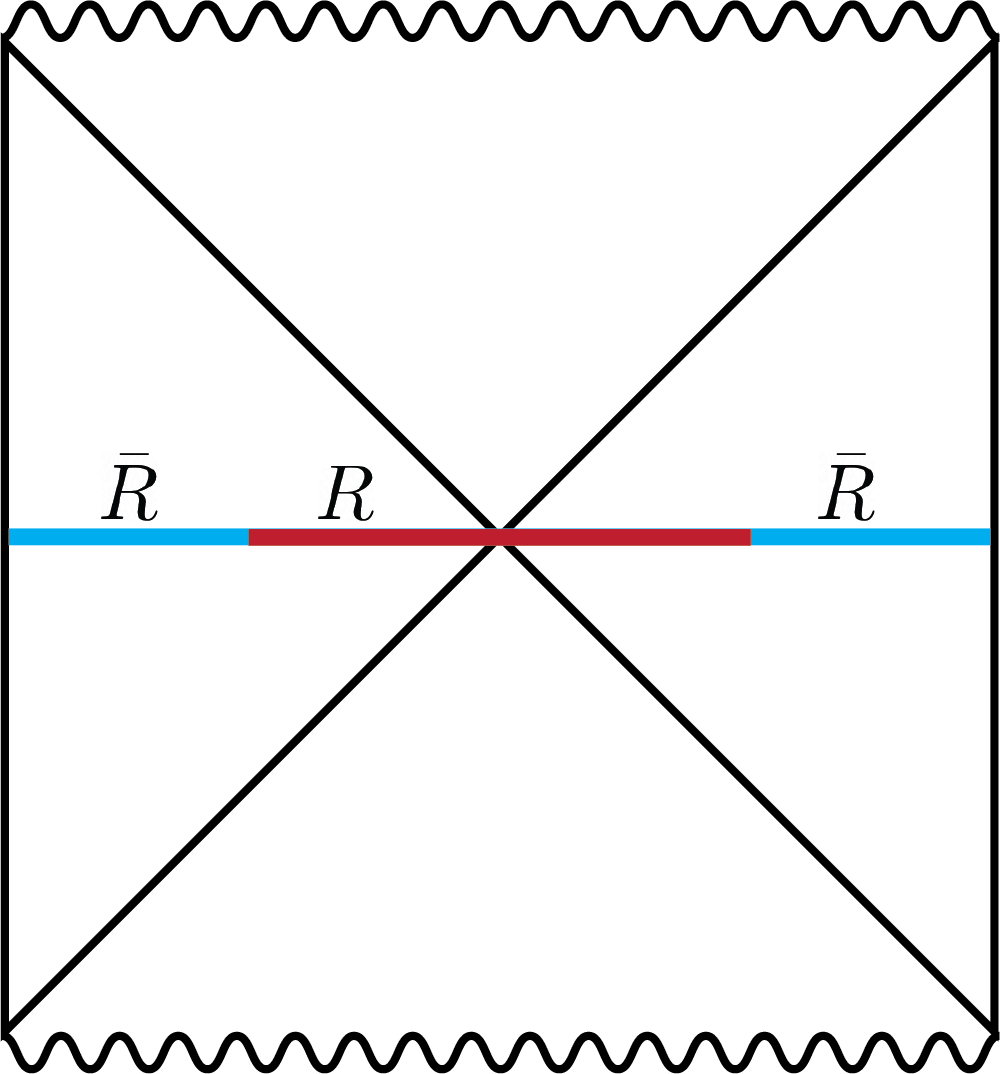}
 \caption{The bulk gravitational system we study in this paper is dual to a BCFT thermal state on a nongravitating black-hole background. We compute the entropy of the subregion $\mathcal{R}$, a connected interval that runs through the \textit{bifurcation surface} (the horizon at $t = 0$ separating the past and future horizons, represented by the point in the middle of the Penrose diagram above), using the holographic RT prescription. This bipartition is similar to that in a recent work \cite{Geng:2021wcq} studying entanglement islands in de Sitter space.
}
\label{fig:penrose}
\end{figure}

\noindent The transparent boundary conditions are responsible for an important and sometimes overlooked result, which is that the graviton on the KR brane's gravity theory picks up a mass \cite{Porrati:2001gx,Porrati:2002dt,Porrati:2003sa} from energy exchange. In the dual ``boundary" description, the transparent boundary conditions leads to the nonconservation of the stress tensor at the BCFT defect, which causes the stress tensor to pick up an anomalous dimension corresponding to a mass for the graviton \cite{Aharony:2006hz}. The graviton's mass is roughly quadratic in brane angle for small angles (the ``near-critical" regime) \cite{Miemiec:2000eq,Schwartz:2000ip} and comparable to the AdS$_{d+1}$ length scale when the angles are not parametrically small.

As discussed in \cite{Geng:2020qvw}, all reliable calculations of the semiclassical Page curve in more than $1+1$ spacetime dimensions ($d > 2$) have been performed in systems with massive gravitons. It was further explained in \cite{Geng:2021hlu} that this mass plays an essential role in enabling islands to constitute a consistent entanglement wedge. Consistent with these statements, the setup in which we study islands in this work features massive gravity on the brane.

\subsection{The Black-String Model} \label{sec2.2}

This section provides an overview of the doubly holographic black string illustrated in Figure \ref{fig:holography}. Starting with a $(d+1)$-dimensional black string, we embed a $d$-dimensional KR brane \cite{Randall:1999vf,Karch:2000ct}. By the AdS/BCFT correspondence \cite{Takayanagi:2011zk,Fujita:2011fp}, the setup is dual to a thermal BCFT state on a black hole background whose degrees of freedom \cite{Affleck:1991tk,Azeyanagi:2007qj} are counted by the brane tension. The intermediate picture is a gravitating black hole coupled to a nongravitating black-hole bath.\footnote{One can also start with empty AdS$_{d+1}$ instead of the black string \cite{Chen:2020uac}. This would produce an analogous picture lacking any black holes.}

The planar black string (with the AdS$_{d+1}$ radius set to $1$) is described by the metric,
\begin{equation}
\begin{split}
ds^2 &= \frac{1}{u^2 \sin^2\mu}\left[-h(u) dt^2 + \frac{du^2}{h(u)} + d\vec{x}^2 + u^2 d\mu^2\right],\\
h(u) &= 1 - \frac{u^{d-1}}{u_h^{d-1}},
\end{split}\label{eq:BS}
\end{equation}
where $t \in \mathbb{R}$, $u > 0$, $0 < \mu < \pi$, and $\vec{x} \in \mathbb{R}^{d-2}$. The blackening factor $h(u)$ gives a black hole on each constant $\mu$ slice---one such slice will be the end-of-the-world KR brane.

We briefly mention that one could also consider the global version, in which case the metric would be,
\begin{equation}
\begin{split}
ds^2 &= \frac{1}{u^2 \sin^2\mu}\left[-h_G(u) dt^2 + \frac{du^2}{h_G(u)} + d\Omega^2_{d-2} + u^2 d\mu^2\right],\\
h_{G}(u) & = 1+u^2-\frac{u^{d-1}}{u_h^{d-1}}. 
\end{split}\label{eq:BSsphere}
\end{equation}
In the global metric, instead of the transverse coordinates comprising $\mathbb{R}^{d-2}$, they form a $(d-2)$-sphere with a line element proportional to $d\Omega_{d-2}$. The dual boundary state is a finite temperature CFT state on a sphere. However, the global black string features a Gregory-Laflamme instability \cite{Gregory:1993vy}. This thermodynamic instability was identified by \cite{Chamblin:2004vr} with a Hawking-Page transition on the brane controlled by a dimensionless combination of the horizon temperature and the boundary sphere's radius. In this paper we will consider \emph{only} the planar black string, for which the Hawking-Page transition is absent.

The brane is introduced by adding the standard Randall-Sundrum (RS) term \cite{Randall:1999vf} to the Einstein action, and then restricting to subcritical brane tensions. The brane is parametrized as $\mu = \theta_b$. The model dual to BCFT is then obtained by treating this as an \textit{end-of-the-world brane} and restricting the $\mu$ coordinate to $\mu \in [\theta_b,\pi)$.\footnote{The tension of the brane is known to be $(d-1)\cos\theta_b$ \cite{Takayanagi:2011zk,Fujita:2011fp}.} 

Our main objective is to use the RT prescription to calculate the entanglement entropy between a BCFT subregion $\mathcal{R}$ and its complement $\bar{\mathcal{R}}$ induced by a generic bipartition of the initial time ($t = 0$) slice, as depicted in Figure \ref{fig:penrose}.
%, which includes the defect dual to the brane. These regions are dual to the region in the bulk containing the black string, and its complement, respectively \cite{Jafferis:2015del,Dong:2016eik}. 
From here the prescription is to determine the minimal extremal surfaces $\Sigma$ that are either homologous to $\mathcal{R}$ or terminate on the end-of-the-world brane \cite{Fujita:2011fp,Almheiri:2019yqk}, with the latter leading to islands on the brane \cite{Penington:2019npb,Almheiri:2019psf}. First we consider a generic interval on the brane $\mathcal{I}$---a ``candidate" island.\footnote{$\Sigma$ simply being homologous to $\mathcal{R}$ corresponds to $\mathcal{I} = \varnothing$, i.e. the empty set.} We then determine extremal surfaces $\Sigma$ that satisfy the homology constraint,
\begin{equation}
\partial\Sigma = \partial \mathcal{R} \cup \partial \mathcal{I}.
\end{equation}
The next step is to apply a variational principle to the \textit{area density functional} $\mathcal{A}$,
\begin{equation}
\mathcal{A} = \int_{\theta_b}^{\pi} \frac{d\mu}{(u\sin\mu)^{d-1}} \sqrt{u^2 + \frac{u'(\mu)^2}{h(u)}}, \label{eq:AF}
\end{equation}
to find the $\Sigma$ and corresponding $\mathcal{I}$ for which the area is minimized. Note that $u_h$ can be brought to the front of the action as an overall prefactor $u_h^{2-d}$ in planar coordinates. For simplicity we choose $u_h=1$ for all of our numerical calculations.

The action \eqref{eq:AF} is extracted from the constant-$t$ slices of the line element $\eqref{eq:BS}$ as the area density of a surface $u = u(\mu)$ with respect to the transverse space $\vec{x}$.\footnote{For the rest of this paper we will use area density and area interchangeably.} This determines the area of the surface up to a factor, proportional to the volume of the suppressed dimensions, that does not affect the phase structure.

\begin{figure}
 \centering
 \includegraphics[width=\linewidth]{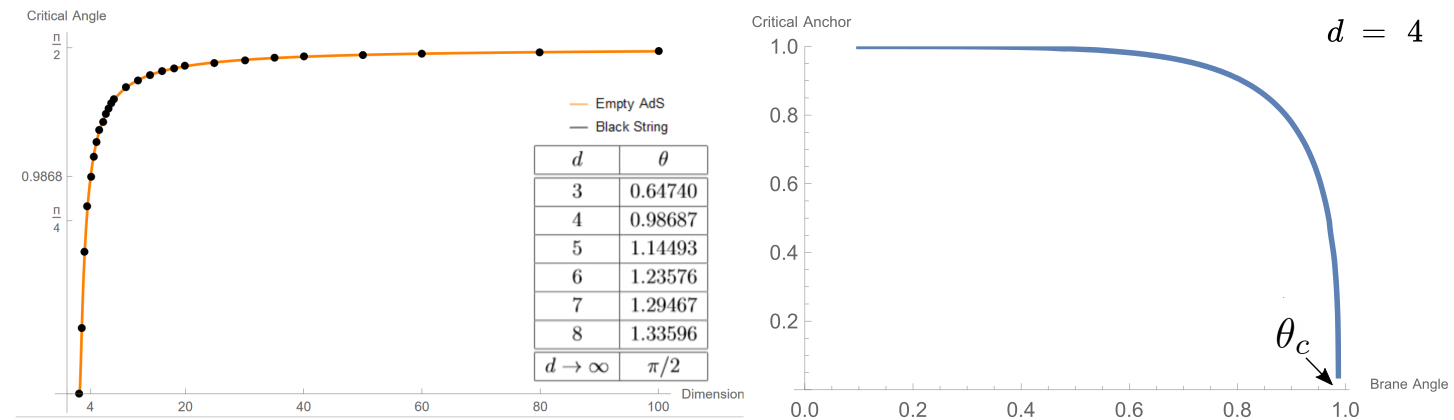}
 \caption{These figures were included in our previous work \cite{Geng:2020fxl}, but we repeat them here for emphasis. The critical angle is a monotonically increasing function of the number of spatial dimensions $d$ and takes the same value for any geometry which is asymptotically AdS$_{d+1}$. This is distinct from the behavior of the critical anchor defining the beginning of the atoll for the black string. The critical anchor instead \textit{decreases} monotonically with brane angle and coincides with the defect at the critical angle (about 0.98687 for $d = 4$). It can be shown that the critical anchor also increases monotonically with the number of dimensions for the space.}
 \label{fig:critical_angle_dim}
\end{figure}

The Euler-Lagrange equation for the action \eqref{eq:AF}, which we also call the ``equation of motion," is an ordinary differential equation because of the parameterization $u = u(\mu)$. Furthermore, we ensure that the boundary terms in the variation of $\mathcal{A}$ vanish by imposing boundary conditions on $\Sigma$. As discussed in \cite{Geng:2020fxl}, we impose a \textit{Dirichlet condition} at the conformal boundary and a \textit{Neumann condition}---requiring that the first derivative $u'$ vanishes---at the brane. The latter tells us that such surfaces must anchor onto the brane at right angles. Solving the Euler-Lagrange equation for the second derivative $u''$ gives,
\begin{equation}
\begin{split}
u'' = -(d-2) u\,h(u) + (d-1) u' \cot\mu \left(1 - \frac{\tan\mu}{2} \frac{u'}{u\,h(u)} + \frac{u'^2}{u^2\,h(u)}\right)
- \left(\frac{d-5}{2}\right) \frac{u'^2}{u}.
\end{split}\label{eomODE}
\end{equation}
The nonlinear first term leads directly to nontrivial physics below the critical angle that we explore in Section \ref{sec2.4}. Figure \ref{fig:critical_angle_dim} reviews how the critical angle depends on the number of dimensions for the space. Note that the critical angle vanishes for $d=2$---the first term in \eqref{eomODE} vanishes and the trajectories of the RT surfaces become circular geodesics that exist for all brane angles.

While we have so far presented the equations for general dimension, we work exclusively with $d=4$ in the remaining sections of this paper.

%%%%%%%%%%%%%%%%%%%%%%%%%%%%%%%%%%%%%%%%%%%%%%%%%%%%%%
\subsection{Different Classes of RT Surfaces}\label{sec2.3}
%%%%%%%%%%%%%%%%%%%%%%%%%%%%%%%%%%%%%%%%%%%%%%%%%%%%%%

\begin{figure}
 \centering
 \includegraphics[width=.7\linewidth]{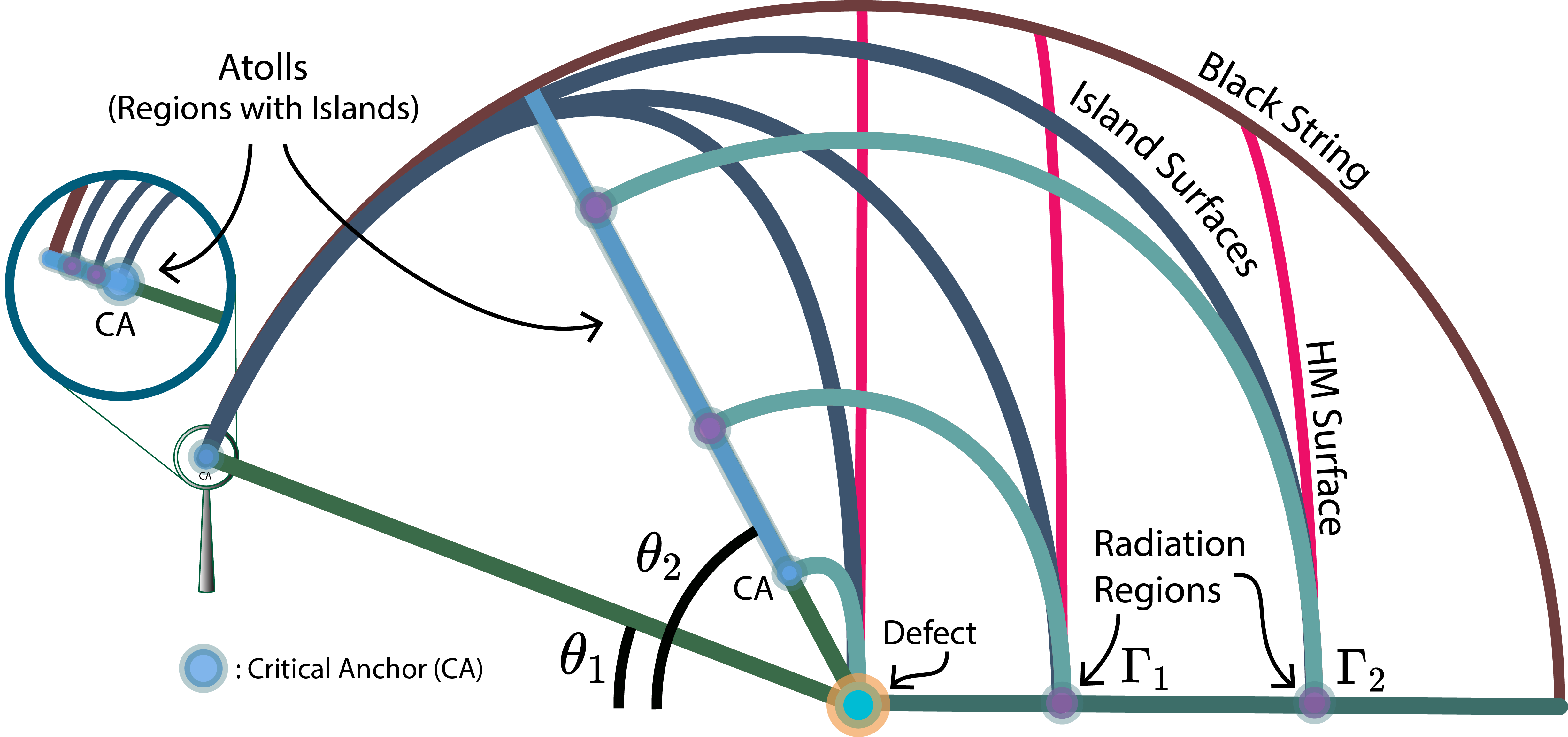}
 \caption{This cartoon illustrates the relevant RT surfaces for different brane angles $\theta_{1,2}$ below the critical angle. Island surfaces end on the brane at right angles, and HM surfaces cross the black-string horizon and traverse the Einstein-Rosen bridge. Islands always attach to the brane inside the atoll, which begins at a point called the \textit{critical anchor} and extends to the black-hole horizon. Decreasing the brane angle reduces the size of the atoll, which shrinks to the horizon as $\theta \rightarrow 0$. Increasing the size of $\mathcal{R}$, bounded by $\Gamma$, pushes the anchors (in purple) toward the black-hole horizon; increasing the brane angle pushes the anchors toward the defect. The critical anchor, which is the point on the brane that determines the size of the largest island on the brane (which always corresponds to $\Gamma=0$%consistent with the Russian doll rule
 ), coincides with the defect above the critical angle. See Figure \ref{fig:surfaces} for numerical results.}
 \label{fig:RT_cartoon}
\end{figure}

We now take a closer look at the Dirichlet and Neumann boundary conditions that play an important role in the construction of the RT surfaces. The relevant surfaces are illustrated in Figure \ref{fig:RT_cartoon}. The two parameters of interest are $\Gamma$, which is the endpoint of the relevant subregion $\mathcal{R}$,\footnote{Technically, there are two endpoints, with one on each side. This subtlety is particularly important for the surfaces explored in Section \ref{subsechm}, but for now it suffices to say that we assume that $\mathcal{R}$ is symmetric with respect to the boundary black hole's bifurcation surface.} and $\theta_b$, the angle between the brane and the excised boundary region. There are two classes of surfaces because, while all surfaces obey Dirichlet boundary conditions at $\Gamma$, only some surfaces connect to the brane where the boundary condition must be Neumann by diffeomorphism-invariance.\footnote{It is mathematically possible to impose a Dirichlet boundary condition on the brane \cite{Ghosh:2021axl}; the Neumann condition is motivated by the presence of localized gravity.}

The Hartman-Maldacena (HM) surfaces avoid the branes as they plunge from the boundary into the black-string horizon, pass through the Einstein-Rosen bridge to the other side, and anchor to the boundary of the thermofield double space. As the Einstein-Rosen bridge grows, the HM surfaces are stretched out behind the horizon, and their areas grow \cite{Hartman:2013qma}.

The counterparts to the HM surfaces are the ``island surfaces" that connect the brane to the conformal boundary. Since these avoid the Einstein-Rosen bridge, their areas are constant in time. Near the conformal boundary they are roughly the same as the HM surfaces at $t = 0$, since they also obey the Dirichlet condition there by virtue of being extremal. However, the island surfaces satisfy a Neumann boundary condition at the brane.

The interplay between these two surfaces makes it possible for the entanglement entropy to obey unitarity. Essentially, island surfaces prevent the entropy from growing forever, since they are constant in time and always dominate (i.e. are minimal) at sufficiently late times. Specifically, even if an island surface is subdominant on the initial time slice to an HM surface in the entanglement entropy, the latter will eventually outgrow the former.

This phase transition of surfaces takes place at the \textit{Page time}. Its value depends on parameters that are clearly connected to the physics of the BCFT state; these are $\theta_b$, which determines the number of degrees of freedom on the defect \cite{Takayanagi:2011zk,Fujita:2011fp}, and $\Gamma$, which determines the size of $\mathcal{R}$. The Page time is roughly proportional to the area difference between the RT surfaces at $t = 0$.

%%%%%%%%%%%%%%%%%
\subsubsection{Hartman-Maldacena (HM) Boundary Conditions}\label{subsechm}
%%%%%%%%%%%%%%%%%

Here we look at the Hartman-Maldacena surfaces in more detail. An alternative $\mu = \mu(u)$ parameterization for the area functional is sometimes useful:
\begin{equation}
\mathcal{A} = \int \frac{du}{(u \sin\mu)^3} \sqrt{\frac{1}{h(u)}+u^2 \mu'(u)^2} \ .\label{eq:AFmu}
\end{equation}
Since the HM surface traverses the Einstein-Rosen bridge, we must specify the location of $\mathcal{R}$ on \emph{both} sides of the geometry. Calling the endpoint of $\mathcal{R}$ and its thermofield double partner $\Gamma$ and $\tilde{\Gamma}$ respectively, for simplicity we consider the symmetric case $\Gamma=\tilde{\Gamma}$. In our previous paper \cite{Geng:2020fxl}, in which we computed ``left/right" entanglement entropy with a \emph{gravitating} bath, we could consider only the case where $\Gamma = 0$. With the nongravitating bath, the choice of $\Gamma$ is no longer restricted.

Choosing the symmetric case leads to a constraint for the HM surfaces on the initial time slice---there will be a ``kink" unless the surfaces cross the black-string horizon orthogonally in tortoise-like coordinates:
\begin{equation}
\label{matchcoef}
 dr = \frac{du}{u \sqrt{h}} \, , \qquad \mu'(r=0)=0 .
\end{equation}
This can be translated into our $\mu(u)$ parameterization by expanding around the horizon:
\begin{equation}
\mu(r) = \sum_{n=0}^{\infty} \mu_{n} r^n \sim \sum_{n=0}^{\infty} \mu_{n} (u_{h}-u)^{n/2} .\label{muExpansion}
\end{equation}
This expansion has to be done in half-integer powers of $(u_h-u)$ because $r \sim \sqrt{u_h-u}$.

The condition that $\mu'(r=0)=0$ in the tortoise-like coordinates implies the absence of a term linear in $r$. Thus the $\sqrt{u_h - u}$ term in the expansion in the $u$ coordinate vanishes, so $\mu(u)$ becomes a regular power series. Imposing the equation of motion in terms of $\mu(r)$ now fixes $\mu'(u_h)$ in terms of $\mu(u_h)$. So even though it naively looks like we imposed only one boundary condition at $u_h$, regularity as implied by the symmetric ansatz tells us that the solution is completely fixed by simply specifying $\mu(u_h)$.

To ensure that we construct an HM surface with a particular value of $\Gamma = u/u_h$, we need to solve the shooting problem with boundary conditions:
\begin{equation}
\mu(u_h \Gamma)=\pi, \hspace{1cm}
\mu(u_{h}) = \theta_{HM}, \hspace{1cm} \mu'(u_{h}) = \frac{2 \cot \theta_{HM}}{u_{h}},\label{eq:bdryHM}
\end{equation}
where $\Gamma$ is the boundary of $\mathcal{R}$ relative to $u_h$, $\theta_{HM}$ is the angle $\mu$ where the HM surface intersects the horizon, and the prime indicates a derivative with respect to the radial coordinate $u$. The Dirichlet condition $\mu'(u_h \Gamma)=\infty$ at the conformal boundary is enforced automatically by the equations of motion. If one prefers the $u(\mu)$ parameterization, one can obtain the same results by inverting the boundary conditions instead.\footnote{This involves the reciprocals of the derivatives in \eqref{eq:bdryHM}. This reparameterization from $\mu(u)$ to $u(\mu)$ is consistent because as we proved in \cite{Geng:2020fxl} that any solution $u(\mu)$ of \eqref{eomODE} is a monotonic function.}

%%%%%%%%%%%%%%%%%
\subsubsection{Island Surface Boundary Conditions}
%%%%%%%%%%%%%%%%%

The island surfaces avoid the Einstein-Rosen bridge and terminate on the brane $\mu = \theta_b$. On the initial time slice they are distinguished from the Hartman-Maldacena surfaces by their resulting brane boundary conditions, which are Neumann and thus require the island surfaces to connect to the brane orthogonally. So, we need to solve the shooting problem for:
\begin{equation}
u(\pi) = u_h\Gamma, \hspace{1cm} 
u(\theta_b) = u_b, \hspace{1cm} u'(\theta_b) = 0,\label{eq:bdryi}
\end{equation}
where $u_b$ indicates the location on the brane where the surface terminates. The condition $u'(\pi)=0$ is enforced automatically by the equations of motion.

Since these surfaces do not pass through the black-string horizon, they do not traverse the Einstein-Rosen bridge and their areas do not change in time. The generic late-time dominance of island surfaces over HM surfaces leads to the saturation of entanglement entropy at the Page time, \emph{provided the island surfaces actually exist}.\footnote{We define the Page time as $t=0$ when the island dominates at the beginning.} But past work \cite{Chen:2020hmv,Geng:2020fxl} has shown these solutions \emph{do not exist} for empty AdS below the critical angle, as we now review.

\subsection{A Potential Puzzle for Unitarity in Higher Dimensions}\label{sec2.4}

Recent work performed independently by our group \cite{Geng:2020fxl}, and others \cite{Chen:2020hmv}, found that island solutions \emph{cease to exist} below a critical angle in empty AdS$_{d+1}$. The angle of the brane, which we reiterate encodes the number of degrees of freedom at the defect and uniquely determines the brane tension in the bulk, plays an important role in determining whether (and where) islands can exist on the brane. This potentially leads to a serious problem for unitarity in all thermal doubly holographic states---without an island surface in the bulk the HM surface would grow without bound.

We note that the previously observed failure of islands to exist occurs when they are not needed to preserve unitarity. Empty AdS$_{d+1}$ with a KR brane is the prime example---RT surfaces are forbidden from traveling between the boundary and any brane below the critical angle, which means the region containing islands on the brane (the atoll) does not exist in this regime. However, since there is no Einstein-Rosen bridge there is no potential paradox involving a monotonically growing entanglement entropy and therefore there is no need for a phase transition between entanglement surfaces in such vacuum states. The nonexistence of island surfaces would, however, lead to a breakdown of unitarity in \textit{thermal} states, such as the black-string geometry considered in this paper.

The preservation of unitarity is thus more subtle for the black-string model because it features an Einstein-Rosen bridge. We find that while island surfaces still exist below the critical angle, they connect only to a finite region of the brane---the atoll. These islands are sufficient to prevent the entanglement entropy from growing without bound, as we showed in a previous paper \cite{Geng:2020fxl} in the context of "left/right" entanglement, in which we are forced to $\Gamma=0$. There we found a critical anchor, which is the boundary of the atoll we find in the current context. In this paper, we find a more complex phase structure associated with the freedom to choose the size of the radiation region on the boundary.

Our results show that for each choice of brane angle below the critical angle, the largest possible island is formed when the island surface begins at the critical anchor and terminates on the defect. This suffices to determine the atolls (see Figure \ref{fig:RT_cartoon}). The critical anchor coincides with the defect above the critical angle, so the atoll completely covers the brane in that case.

%%%%%%%%%%%%%%%%%
\section{Numerical Results for the Area Difference \label{secardiff}}
%%%%%%%%%%%%%%%%%

Here we present our numerical results for the initial area difference between the competing island and HM surfaces as a function of the two parameters $\Gamma$ and $\theta_b$. The intricacy of our results compared to those found with a gravitating bath \cite{Geng:2020fxl} arises from the freedom to vary $\Gamma$ on the nongravitating bath. The phase structure is determined by the sign of the UV-finite area difference,
\begin{equation}
\Delta\mathcal{A}(t) = \mathcal{A}_{IS} - \mathcal{A}_{HM}(t). \label{areaDiff}
\end{equation}
The sign is important because it tells us which type of surface is dominant in the entanglement entropy (i.e. minimal in area). The HM surface is dominant when $\Delta\mathcal{A}(t) > 0$, and the island surface is dominant when $\Delta\mathcal{A}(t) < 0$. To determine whether or not we have a Page curve at all, it is sufficient to look at the $t = 0$ slice, since the HM surface is at its smallest in $t$ when it crosses the bifurcation surface of the black string.

\begin{figure}
 \centering
 \includegraphics[width=\linewidth]{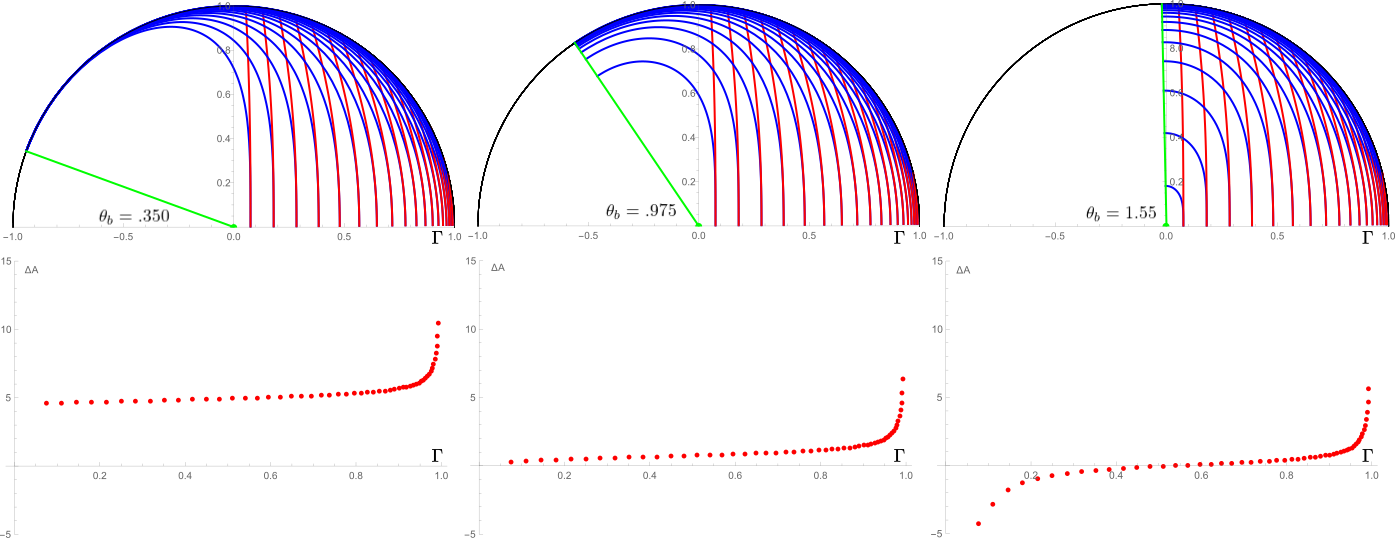}
 \caption{Here we see the setup together with numerical data in the planar black string. The KR brane is green, the HM surfaces are red, the island surfaces are blue, and the black-string horizon is the semicircle in black. The area difference between the island surface and the HM surface for various anchor points $\Gamma$ is displayed in the set of second plots below. Sending $\Gamma \rightarrow 0$ recovers the finite area differences from our previous paper \cite{Geng:2020fxl} for angles below the critical angle. When the brane lies above the critical angle, the area difference diverges to negative infinity. These divergences are explained physically in Section \ref{secdiv}.}% reflecting the transition from finite island surfaces to tiny island surfaces. Since we don't explicitly mention the tiny islands in this paper, I think this comment would just confuse the reader. If needed we can add a footnote. Mark}
\label{fig:surfaces}
\end{figure}

Through the boundary conditions \eqref{eq:bdryHM}--\eqref{eq:bdryi}, the initial area difference $\Delta\mathcal{A}(0)$ depends on just two free parameters, the anchor point on the bath $\Gamma$ and the angle of the brane $\theta_b$. We mentioned before that $u_h$ scales out of the action, so we are free to choose $u_h=1$ for our numerical results. We find that $\Delta\mathcal{A}(0)$ is finite when the parameters lie within the following open intervals,
\begin{equation}
\Gamma \in (0, 1), \qquad \theta_b \in (0, \pi).
\end{equation}
See Figure \ref{fig:surfaces} for samples of our numerical data at brane angles below, at, and above the critical angle. The question of what happens when one parameter is fixed while the other varies is explored in Figures \ref{fig:fixed_anchor_points} and \ref{fig:fixed_brane_angles}. $\Delta\mathcal{A}(0)$ as a function of both $\Gamma$ and $\theta_b$ is plotted in 3D in Figure \ref{fig:symmetry}.

%%%%%%%%%%%%%%%%%
\subsection{The Difference between the Page Angle and Critical Angle}
%%%%%%%%%%%%%%%%%

There are two special angles that play an important role in this section---the \textit{Page angle} $\theta_P$ and the \textit{critical angle} $\theta_c$ \cite{Geng:2020fxl}. For clarity, we briefly review these before presenting our numerical results. 

The Page angle is the value of $\theta_b$ at which $\Delta\mathcal{A}(0)$ (for $\Gamma = 0$) vanishes and the entanglement entropy becomes constant, and the critical angle is the slightly larger value of $\theta_b$ below which extremal island surfaces do not exist in empty AdS$_{d+1}$. Note however that the critical angle still plays a role in other geometries so long as the asymptotic region is still empty AdS$_{d+1}$. In particular, when the brane is below the critical angle, island surfaces approach the horizon but do not probe the asymptotic region. The islands on the braneworld are restricted to the atoll discussed in Section \ref{sec2.4}, which is only a finite portion of the brane below the critical angle, whereas above the critical angle the atoll covers the entire brane.

In empty AdS, the Page angle and the critical angle are equal, but in other geometries they are generally slightly different. The area difference vanishes slightly below the critical angle in the black string geometry because the area of the HM surface is bigger than it would be in empty AdS, but the island solution for $\Gamma=0$ is unchanged at the critical angle.

%%%%%%%%%%%%%%%%%
\subsection{Numerical Results for Fixed Parameters}
%%%%%%%%%%%%%%%%%

Here we take a closer look at the initial area differences in our numerical results, which were computed for thousands of combinations of the anchor point $\Gamma$ and the brane angle $\theta_b$. We depict a representative portion of the full $(\Gamma,\theta_b)$ parameter space.

\begin{figure}
 \centering
 \includegraphics[width=.8\linewidth]{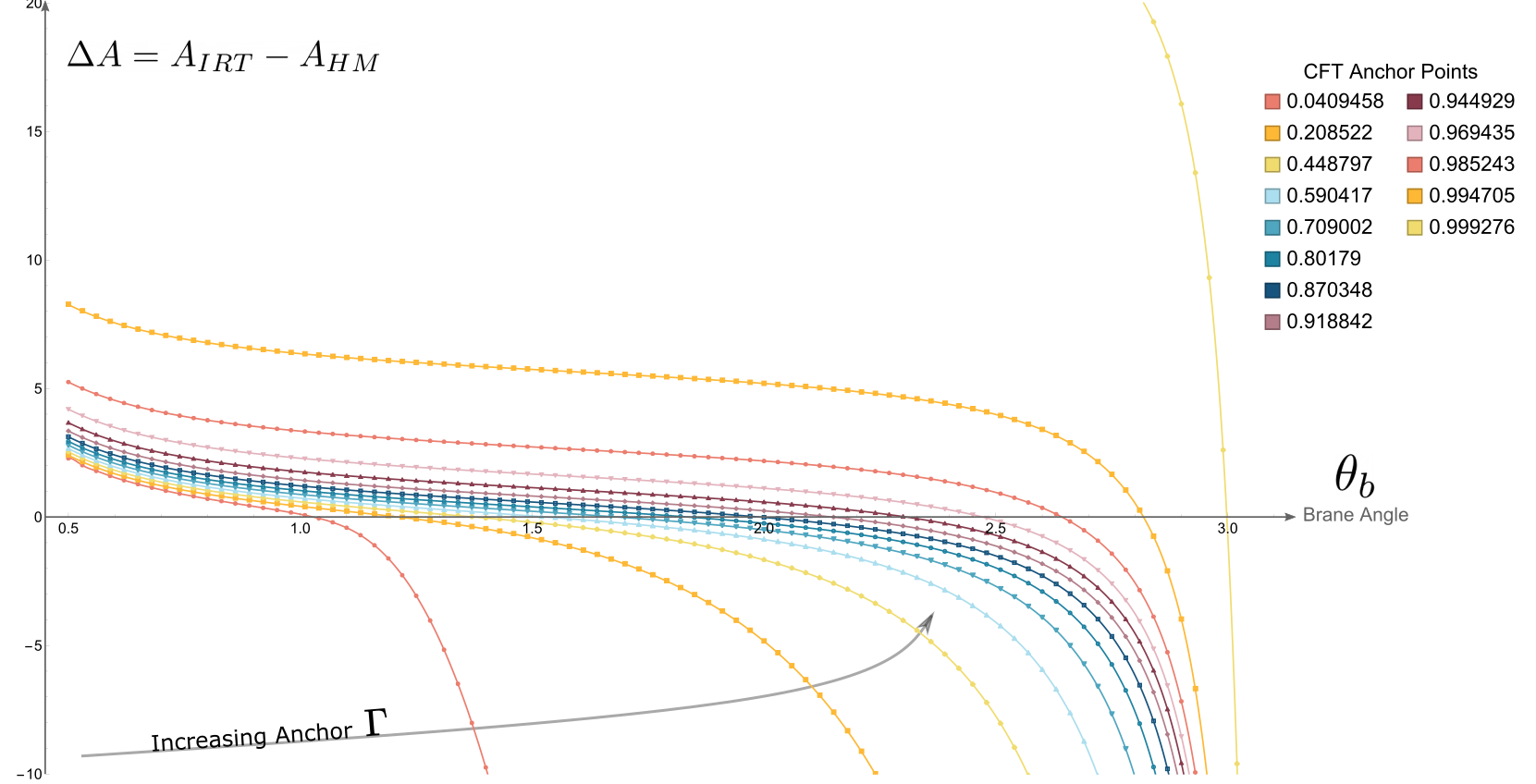}
 \caption{ The area difference is displayed as a function of brane angle $\theta_b$ for many possible CFT anchor points $\Gamma$. The area difference is clearly monotonically increasing in $\Gamma$ and monotonically decreasing in $\theta_b$. The anchor points were determined numerically for many values of $\theta_{HM}$ using the shooting method with the boundary conditions in \eqref{eq:bdryHM}. The area difference diverges to negative infinity as $\Gamma \rightarrow 0$ for $\theta_b>\theta_c$, but when $\Gamma\rightarrow 0$ for $\theta_b<\theta_c$ it converges to a finite value.}
 \label{fig:fixed_anchor_points}
\end{figure}

The $t=0$ area differences\footnote{By area differences, we always mean at $t=0$.} ($\Delta\mathcal{A}=\mathcal{A}_{HM}-\mathcal{A}_{island}$) for fixed $\Gamma$ as a function of $\theta_b$ are illustrated numerically in Figure \ref{fig:fixed_anchor_points}. The area difference decreases monotonically with the value of $\theta_b$, and from comparing the different curves on constant-$\theta_b$ slices it can be seen that the area difference increases monotonically with anchor point $\Gamma$. The area difference is negative---and thus the island surface is initially dominant---only when the brane lies above the Page angle.\footnote{Recall that the brane angle is measured from the excised boundary region, see Figure \ref{fig:holography}.} Above the critical angle the area difference diverges at the edges of the intervals $\Gamma \in (0, 1)$ and $\theta_b \in (0, \pi)$, while below the critical angle the area difference is finite as $\Gamma\rightarrow0$. For all other values of $\Gamma$ and $\theta_b$ the area difference is finite. 

\begin{figure}
 \centering
 \includegraphics[width=.8\linewidth]{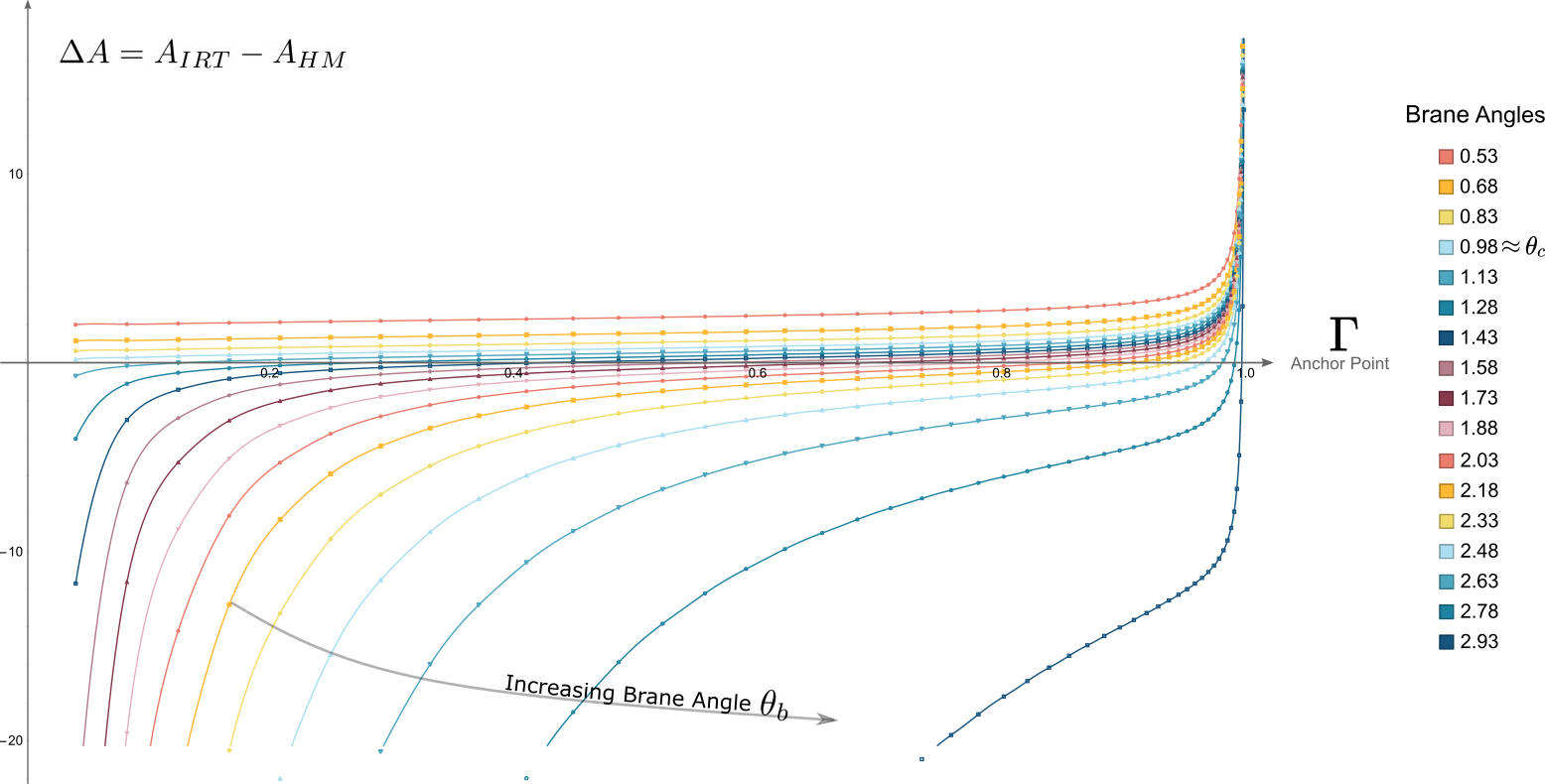}
 \caption{ The area difference is displayed as a function of CFT anchor point $\Gamma$ for many possible brane angles $\theta_b$. The area difference is clearly monotonically increasing in $\Gamma$ and monotonically decreasing in $\theta_b$. The islands become dominant when the area difference is negative, which is possible only above the Page angle. The Page angle \cite{Geng:2020fxl} is very slightly less than the critical angle ($\theta_c \approx 0.98687$ in $d=4$). The area difference diverges to negative infinity as $\Gamma \rightarrow 0$ for $\theta_b>\theta_c$, but when $\Gamma\rightarrow 0$ for $\theta_b<\theta_c$ it converges to a finite value. We explain in Section \ref{secdiv} that the area difference diverges to positive infinity as $\theta_b \rightarrow 0$.}
 \label{fig:fixed_brane_angles}
\end{figure}

Similar considerations hold for fixed brane angles $\theta_b$ as a function of anchor point $\Gamma$, illustrated numerically in Figure \ref{fig:fixed_brane_angles}. Again we see that the area difference increases monotonically with the value of $\Gamma$. From comparing the different curves on constant-$\Gamma$ slices, one can see the area difference is also monotonically decreasing in brane angle $\theta_b$. Again the area difference is negative only when the brane lies above the Page angle.

Note that numerical results are included in the figures only above $\theta_b = 0.5$. This is because the computation time for the most stable\footnote{In order to validate our numerics, we have repeated the calculations of the area difference for a large range of different cutoffs. As we show in Appendix \ref{app:B}, our results are very stable.} version of the shooting method becomes very large as $\theta_b$ becomes small. As a check, the calculation was also performed using a slightly different method for smaller brane angles. While this is not visible in the figures, we found that the area difference diverges to positive infinity as $\theta_b \rightarrow 0$. These numerical results are found to match our earlier work \cite{Geng:2020fxl} in the limit where $\Gamma \rightarrow 0$.

\subsection{The Geometric Origin of the Divergences}\label{secdiv}

It is simple to understand the diverging area difference for parameters at the edges of the intervals $\Gamma\in(0,1)$ and $\theta_b\in(0,\pi)$ geometrically. We reiterate that the surfaces themselves have been illustrated in Figure \ref{fig:surfaces}, the area differences for fixed values of $\Gamma$ and $\theta_b$ have been given in Figures \ref{fig:fixed_anchor_points} and \ref{fig:fixed_brane_angles}. The positive area differences are presented in 3D in Figure \ref{fig:symmetry}.

\begin{figure}
 \centering
 \includegraphics[width=\linewidth]{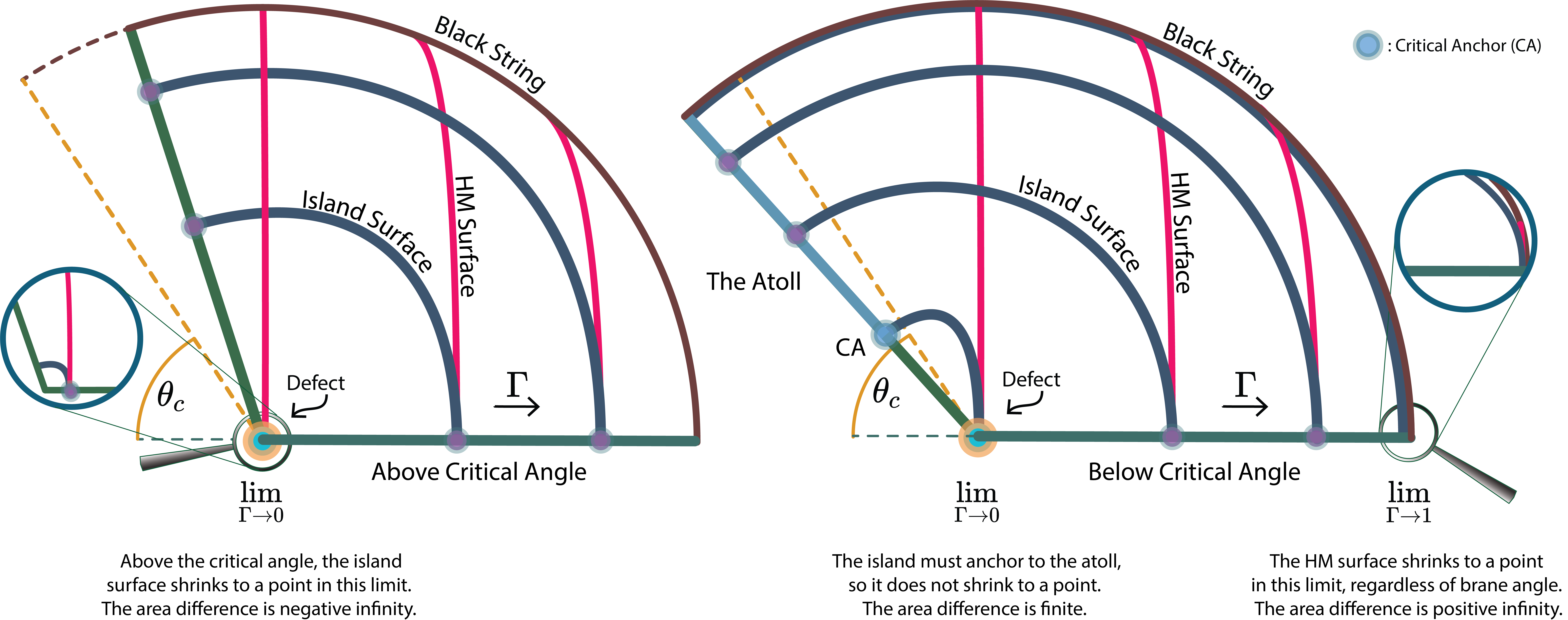}
 \caption{This cartoon illustrates how the area difference is affected by each limiting case for the boundary anchor point $\Gamma$, with angles below the critical angle on the right, and angles above the critical angle on the left. Taking the difference between the surfaces essentially acts as a renormalization scheme. The area difference between the island surface and HM surface diverges when one of them shrinks to a point because the difference between the surfaces reaches farther into the boundary region. When the brane lies below the critical angle, the island surface cannot shrink to a point when $\Gamma\rightarrow0$ because it anchors to the brane inside the atoll.}
 \label{fig:gamma_limit_picture}
\end{figure}

\subsubsection{Positive Divergence}

First we consider each case where the area difference diverges to positive infinity, meaning the island surface has infinitely bigger area than the HM surface. This happens when $\theta_b\rightarrow0$ for each fixed $\Gamma$ and when $\Gamma\rightarrow1$ for each fixed $\theta_b$. The former can be understood by counting how many times each RT surface hits the conformal boundary; in a single exterior patch of the black string, the island surface touches the boundary twice while the HM surface touches it only once. The latter case where $\Gamma\rightarrow1$ is illustrated in Figure \ref{fig:gamma_limit_picture}. Since the HM surface shrinks to a point on the boundary, it only serves to ``cut off the end" of the island surface, so the area difference diverges to positive infinity. 

The divergence can be understood as follows. As we approach the limiting value $\Gamma = 1$, we are effectively computing the entropy of a small strip on the boundary (see Figure \ref{fig:thin_strip} in Appendix \ref{app:A}). In this limit, the curvature of the boundary black hole is not important so the HM surface reduces to the familiar RT surface for a small strip in empty AdS. It is well known that the renormalized entanglement entropy of a small strip---and, correspondingly, the renormalized area of the HM surface---tends to $-\infty$ in this limit.\footnote{See the second term in \eqref{eq:TakaS} in Appendix \ref{app:A}; this term is the renormalized entanglement entropy}. Therefore the area difference between the area of the island surface and the HM diverges.

\subsubsection{Negative Divergence}

Infinite \emph{negative} area differences do not lead to eternally-increasing entropy, since island surfaces start out dominant on the initial time-slice for \emph{every} negative area difference. The HM surface starts out infinitely larger than the island surface, so the entropy is simply constant and lacks a phase transition. But we can still consider the cases leading to such divergences.

The area difference diverges to negative infinity whenever the island surface shrinks to a point, as illustrated in Figure \ref{fig:gamma_limit_picture}; this takes place when $\Gamma\rightarrow0$ for each fixed $\theta_b$ above the critical angle $\theta_c$. Note that these are the ``tiny" island surfaces found in \cite{Geng:2020fxl}.

There are also some nonphysical limits which give diverging negative area differences, for example in the limit where $\theta_b \rightarrow \pi$, but note the brane tensions are negative when $\theta_b > \pi/2$. For example, the area difference also diverges to negative infinity when $\theta_b \rightarrow \pi$ for a fixed $\Gamma$, with the caveat that the associated HM surface takes a nonphysical path \emph{through} the brane and into the excised region.\footnote{Such HM surfaces do not satisfy the homology constraint in any case in the bulk spacetime cutoff by the KR brane.} Fortunately, it turns out that the HM surfaces never travel through the brane when the area difference is positive,\footnote{This was also noted in \cite{Geng:2021iyq} in a lower dimensional setup together with a precise field theory calculation.} and the critical angle \emph{never} takes a value greater than $\pi/2$ for any dimension.

We have seen what happens when $\theta_b \rightarrow \pi$ and when $\Gamma \rightarrow 1$ separately, but it is interesting to note what happens if both limits are applied simultaneously. This is a sensible procedure because you can keep $\theta_{HM}>\theta_{b}$ while taking the limit. It turns out that the area difference vanishes, which can be seen geometrically because the boundary conditions for the HM surface and island surface become degenerate. This can be seen in \eqref{eq:bdryHM}; when the parameterization is switched over to $u(\mu)$, the boundary condition for the HM surface becomes,
\begin{equation}
u'(\mu)|_{u=u_h} = \frac{u_h}{2 \cot \theta_{HM}}. \label{eq:bdryHMu}
\end{equation}
Recalling that $\theta_{HM}$ is the polar angle $\mu$ where the HM surface crosses the black-string horizon, and considering the geometry of the planar black string, we see that $\theta_{HM} \rightarrow \pi$ is the same limit as $\Gamma \rightarrow 1$. When this limit is taken, the boundary conditions in \eqref{eq:bdryi} and \eqref{eq:bdryHMu} match, the surfaces are the same, and the area difference vanishes.

\subsection{Symmetry in Brane Angle and HM Angle}\label{sec:symmetry}

\begin{figure}
 \centering
 \includegraphics[width=\linewidth]{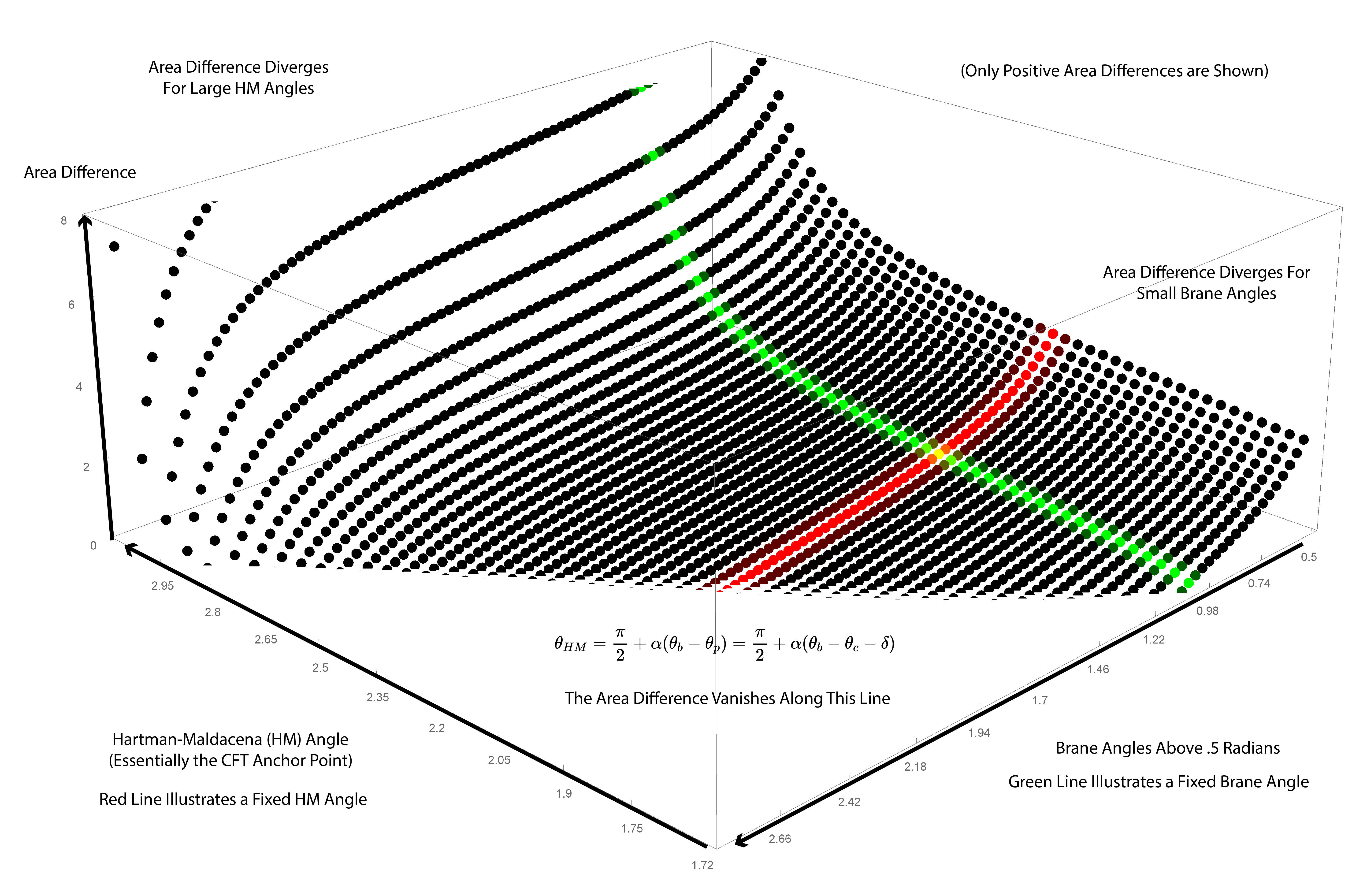}
 \caption{Here we present the positive area difference as a function of $\theta_b$ and $\theta_{HM}$. The parameter subspace for which the area difference vanishes is an approximately straight line. Negative brane tensions are most likely not physical but they have been included without much trouble. This figure is more complete than Figures \ref{fig:fixed_anchor_points} and \ref{fig:fixed_brane_angles}, which simply correspond to slices along the $x$ and $y$ axes, with the main difference being that $\Gamma$ has been swapped out for $\theta_{HM}$. Note the area difference increases monotonically with $\theta_{HM}$ and decreases monotonically with $\theta_b$, and that the area difference vanishes along an approximately straight line.}
 \label{fig:symmetry}
\end{figure}

We observe a striking symmetry between Figure \ref{fig:fixed_anchor_points} and Figure \ref{fig:fixed_brane_angles} which can be made more apparent by replacing $\Gamma$ with the HM angle $\theta_{HM}$, which we recall is the polar angle where the HM surface strikes the event horizon. This redefinition is not as arbitrary as it may seem at first glance. Since the boundary condition is given in polar coordinates, the coordinate transformation is natural from the perspective of solving the shooting problem in \eqref{eq:bdryHM}.

From looking at Figure \ref{fig:symmetry}, one can see the area difference is roughly symmetric in brane angle and HM angle; this can be made more apparent by including $\theta_b < 0.5$ but at the cost of some accuracy. Only positive tension branes are considered on the left, but negative tension branes have been included for completeness on the right.

The symmetry can be used to write an approximately linear constraint on the region of parameter space for which the area difference is positive, with $\Delta\mathcal{A}(0) = 0$ at equality,
\begin{equation}
 \theta_{HM}\ge \frac{\pi}{2} + \alpha(\theta_b - \theta_c - \delta) \iff \Delta \mathcal{A}(0) \ge 0,
 \label{eq:EntropyBelt}
\end{equation}
where $\theta_{c}$ is the critical angle, $\theta_b$ is the brane angle, $\delta \equiv \theta_P - \theta_c$ ($\delta \approx -0.02$ in $d = 4$), and $\alpha \approx 0.75$ in $d=4$. The value of $\alpha$ generically depends on $d$; it equals unity for $d=2$ and vanishes as $d\rightarrow \infty$.

Choosing $\Gamma=0$ recovers the result of \cite{Geng:2020fxl}, since the area difference vanishes at the Page angle $\theta_P = \theta_c + \delta$ for $\Gamma=0$. Furthermore, the area difference vanishes when $\theta_{HM}=\theta_b=\pi$; in that case it can be seen that the surfaces are the same, since the HM and island surface boundary conditions become degenerate in that limit.

It can be seen in this equation that, as we mentioned before, the area difference is always positive when $\theta_{HM}\ge\theta_b$. This serves as a useful consistency check, since the HM surface never travels through the brane for positive area differences, and the area difference vanishes when $\theta_{HM} = \theta_b = \pi$.

%%%%%%%%%%%%%%%%%
\subsection{Phase Structure and the Constant Entropy Belt}
%%%%%%%%%%%%%%%%%

\begin{figure}
 \centering
 \includegraphics[width=.9\linewidth]{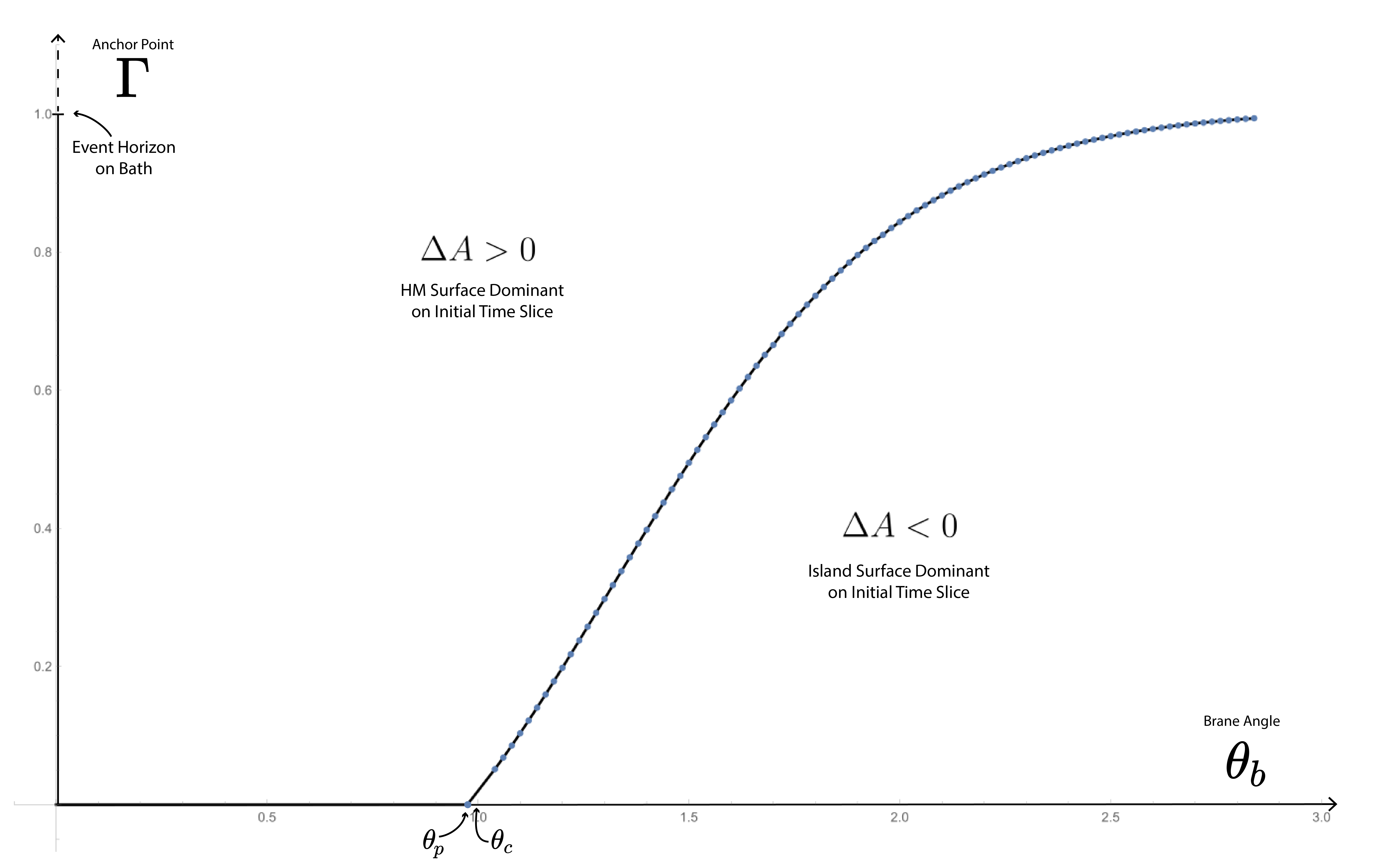}
 \caption{The phase diagram depends on the anchor point $\Gamma$ and the brane angle $\theta_b$. When the brane lies above the Page angle, the island surfaces are dominant when $\Gamma$ is sufficiently small; as the brane angle increases the island surfaces are dominant for increasingly large values of $\Gamma$. For anchor points outside the boundary black hole's horizon ($\Gamma < 1$), the island surfaces become dominant as we take $\theta_b \rightarrow \pi$.}
 \label{fig:phaseDiag}
\end{figure}

Here we explain our main result, the phase structure for the black-string model with a nongravitating bath as a function of the anchor point $\Gamma$ and the brane angle $\theta_b$. Figure \ref{fig:phaseDiag} illustrates the phase transition that occurs for a given $\theta_b > \theta_P$ when $\Gamma$, which separates $\bar{\mathcal{R}}$ from $\mathcal{R}$ in Figure \ref{fig:penrose}, becomes large enough for the entropy to be described by a time-dependent Page curve. When discussing Figures \ref{fig:fixed_anchor_points} and \ref{fig:fixed_brane_angles} we pointed out that the area difference is monotonically decreasing in $\theta_b$ and monotonically increasing in $\Gamma$. This behavior manifests in the phase diagram.

\begin{figure}
    \centering
    \includegraphics[width=.9\linewidth]{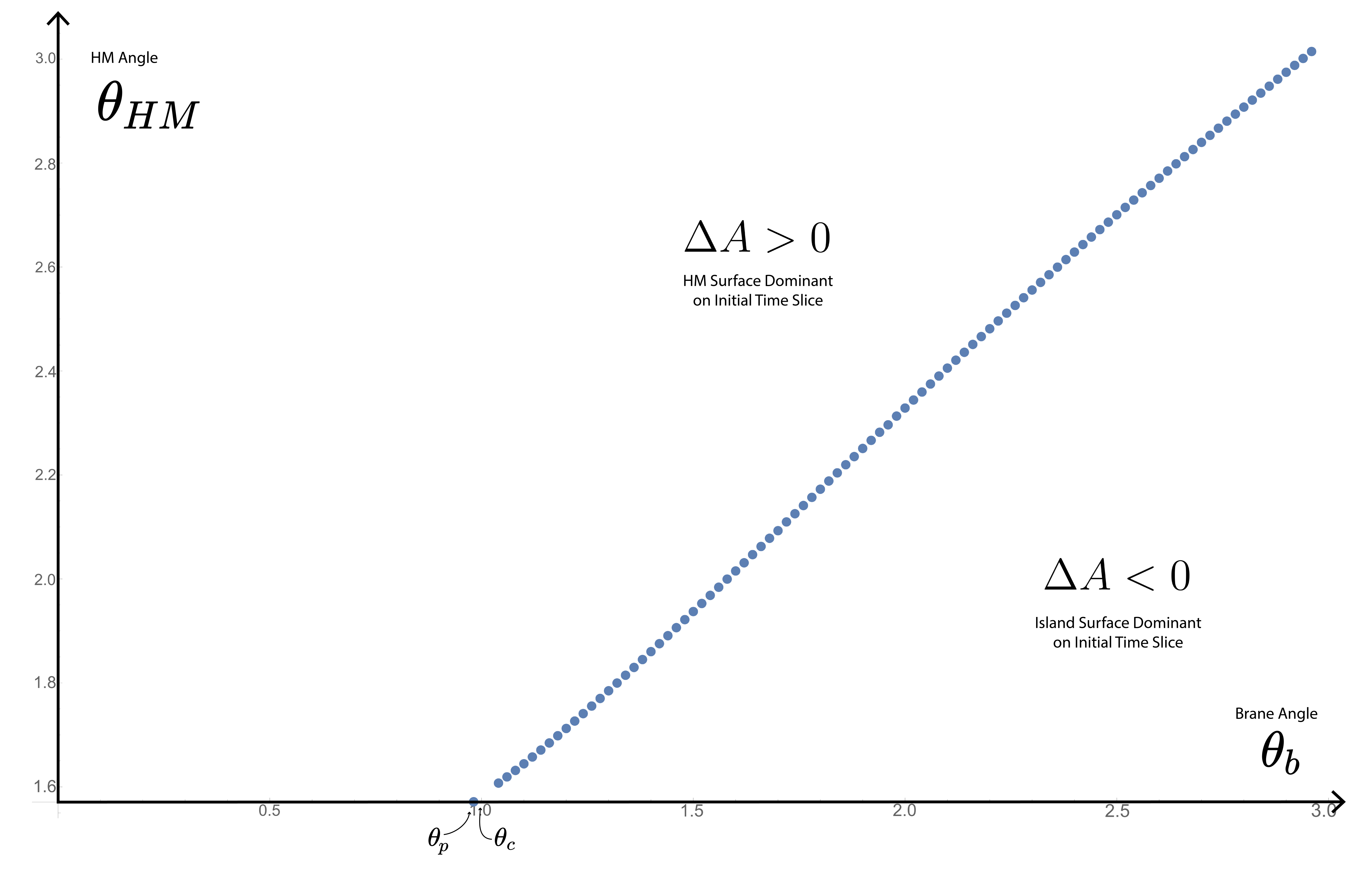}
    \caption{The phase diagram is approximately linear when the anchor point $\Gamma$ is exchanged for the HM angle $\theta_{HM}$; this can also be seen in Figure \ref{fig:symmetry}. The phase diagram is otherwise the same as in Figure \ref{fig:phaseDiag}, with $\theta_{HM}=\pi$ corresponding to $\Gamma=1$ and $\theta_{HM}=\pi/2$ corresponding to $\Gamma=0$.}
    \label{fig:linear}
\end{figure}

It can be seen in Figure \ref{fig:symmetry} and Figure \ref{fig:phaseDiag} that for a given value of $\Gamma$ the initial area difference is negative (thus the island surface is dominant) if $\theta_b$ is large enough. This is precisely the condition \eqref{eq:EntropyBelt}. In such cases the entanglement entropy does not grow with time, since the island surface does not traverse the Einstein-Rosen bridge. In the limit $\Gamma\rightarrow0$, or equivalently $\theta_{HM}\rightarrow \pi/2$, the area difference vanishes at the Page angle \cite{Geng:2020fxl}.

\begin{figure}
 \centering
 \includegraphics[width=.85\linewidth]{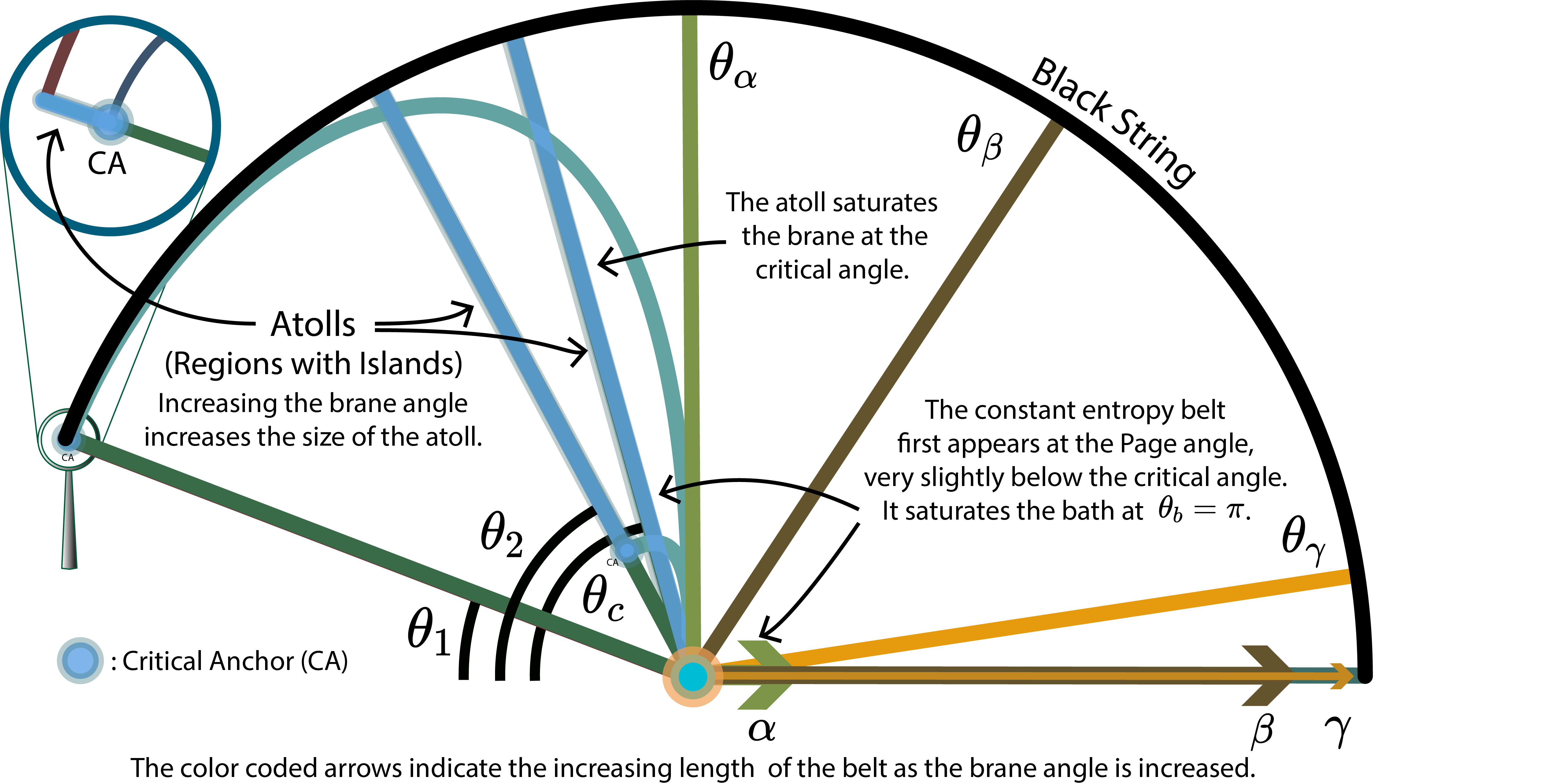}
 \caption{When an RT surface anchors within the constant entropy belt on the bath, the island surfaces are dominant on the initial time slice and the entanglement entropy is constant in time. Decreasing the brane tension $(d-1)\cos\theta_b$ increases the width of the belt, indicated by the color-coded arrows on the boundary which correspond to the vertical axis in Figure \ref{fig:phaseDiag}. The belt first appears at the Page angle, slightly below the critical angle, and covers the entire region outside the boundary black hole horizon (at $\Gamma=1$) in the limit where $\theta_b \rightarrow 1$. All angles are measured from the boundary, and the tensionless limit is achieved at $\theta_b = \pi/2$.}
 \label{fig:entropy_belt}
\end{figure}

The above properties make it possible to nearly summarize the phase structure in terms of what we call the \textit{constant entropy belt}, illustrated in Figure \ref{fig:entropy_belt}. The definition is that the entropy curve is constant, for a given $\theta_b$, when $\Gamma$ lies within the constant entropy belt. In all such cases the area difference is negative, which means the island surface will be dominant on the initial time slice, and therefore the entanglement entropy will be constant in time.

The belt does not exist when $\theta_b$ is below the Page angle because the area difference is positive for all $\Gamma$. The belt first forms at the Page angle, only very slightly below the critical angle, so there is a small parametric range $\theta_b \in (\theta_P,\theta_c)$ for which we have both a constant entropy belt and a finite critical anchor and hence a nontrivial atoll. Whereas the entropy belt (on the boundary) always includes the defect, the atoll (on the brane) extends almost to the defect at the Page angle at which the constant entropy belt first appears. Since the area difference decreases monotonically with brane angle, increasing $\theta_b$ increases the range of $\Gamma$ values for which the entropy is constant, so the belt grows with angle and is always connected. The belt completely covers the region on the boundary outside the boundary black hole's horizon, which is located at $\Gamma=1$, in the limit where $\theta_b \rightarrow \pi$. 

\begin{figure}
 \centering
 \includegraphics[width=\linewidth]{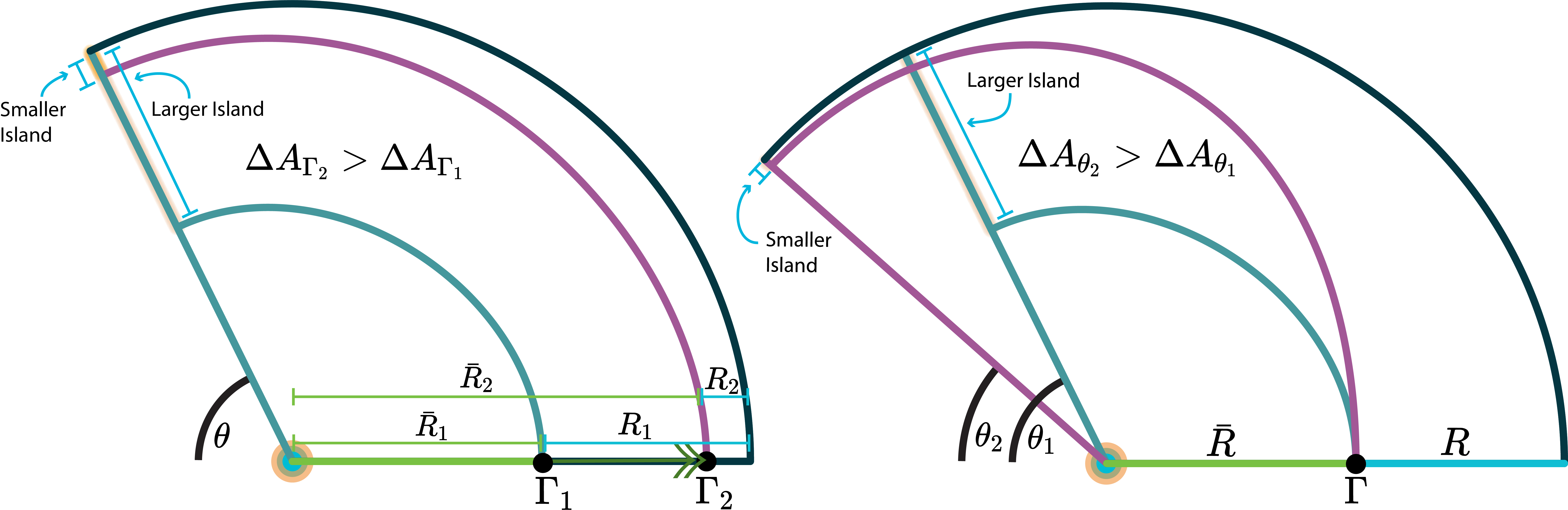}
 \caption{Here we illustrate the \emph{Russian doll rule}, which states that nested surfaces can be constructed by either decreasing the brane angle $\theta_b$ or increasing the anchor point $\Gamma$. This always pushes the surfaces closer to the event horizon into the bulk, decreases the size of the island on the brane, and increases the area difference between the island surface and the HM surface on the initial time slice. The area difference increases without bound as the event horizon is approached by the island surface. }
 \label{fig:russiandoll}
\end{figure}

Figure \ref{fig:russiandoll} illustrates one more noteworthy property of the phase structure; the initial area difference is geometrically related to the size of the island on the brane. Decreasing $\theta_b$ or increasing $\Gamma$ pushes the island surfaces deeper into the bulk and reduces the size of the island. The aforementioned monotonicity in those parameters leads to what we call the \emph{Russian doll rule}, which states that the area difference always increases when $\Gamma$ is increased, which always pushes the island surfaces closer to the horizon in the bulk. This area difference diverges to positive infinity as the black-string horizon is approached by the island surfaces. We expect that the Page time also diverges in this limit.

%For fixed $\Gamma$ Cixin loses access to the island region at the Page time -- however Cixin (as with Bindu) can increase $\Gamma$ to access the island region. 

\subsection{Consequences for Entanglement Wedge Reconstruction \label{subsecentwedgerecon}}

Armed with many copies of this doubly holographic model, experimenters on the boundary could go about reconstructing the bulk region by making many measurements of observables on the boundary region \cite{Jafferis:2015del,Faulkner:2017vdd,Almheiri:2019yqk}. 
%Since the black string extends to the boundary region, 
We suppose %for physical reasons 
that we have two sets of experimenters who can access specific boundary degrees of freedom outside the horizon. For simplicity we will call these groups Cixin and Bindu.

\begin{figure}
    \centering
    \includegraphics[width=\linewidth]{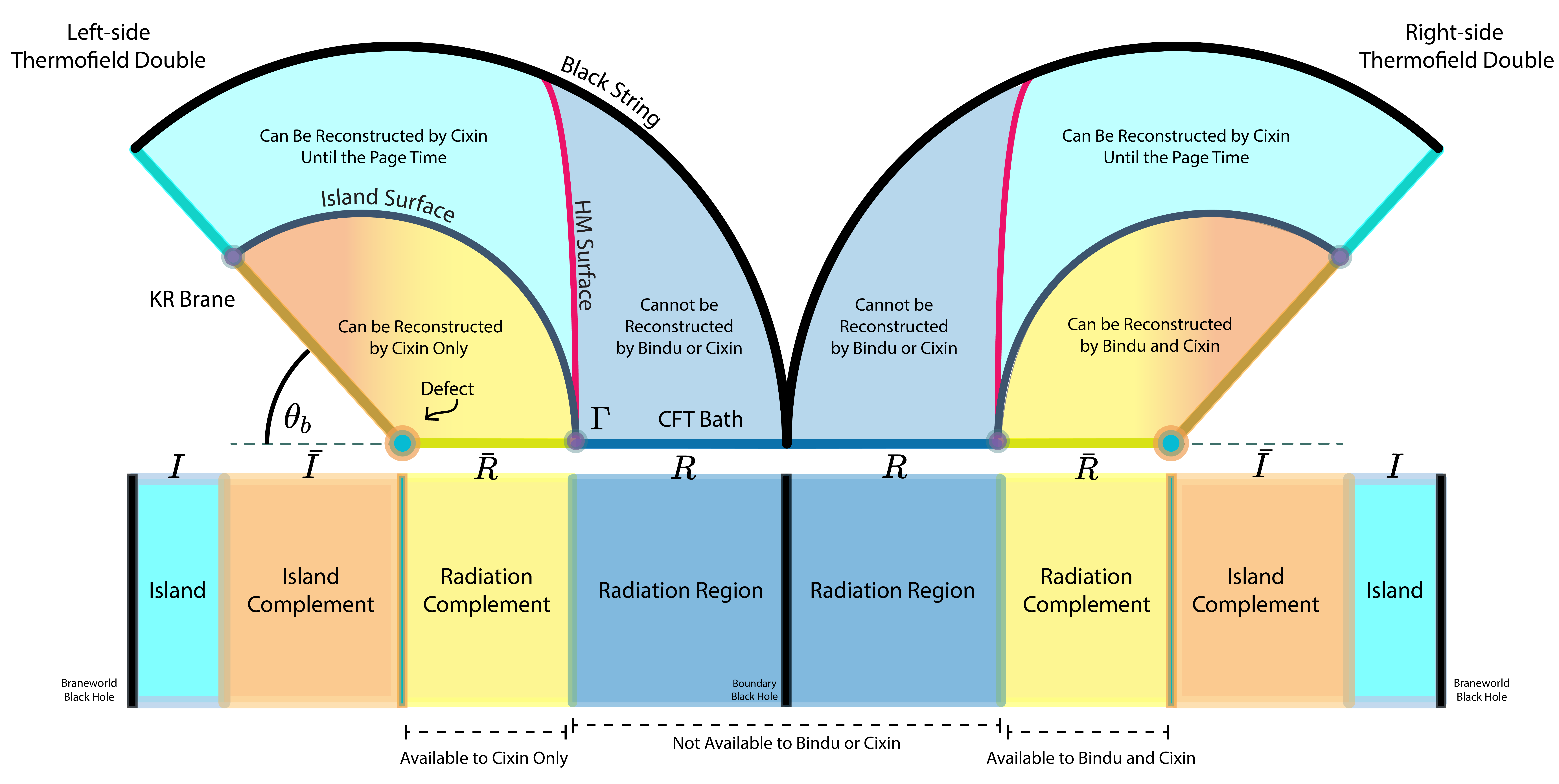}
    \caption{The island region can be reconstructed using boundary data in $\bar{\mathcal{R}}$ only if both sides are accessible and only until the Page time. According to the Russian doll rule, the size of the region which is lost at the Page time increases monotonically with the brane angle $\theta_b$ and decreases monotonically with $\Gamma$.}
    \label{fig:reconstruction}
\end{figure}

The first group (Cixin) has access to boundary degrees of freedom on both sides of the Einstein-Rosen bridge shown in Figure \ref{fig:penrose}, namely the open interval $(0,\Gamma)$ and its thermofield double partner $(0,\tilde{\Gamma})$. The other group (Bindu) has access to only one side, which may as well be the right side, since $\Gamma = \tilde{\Gamma}$ in our model.\footnote{We could have instead considered a class of observers who have access to one side of the radiation region ($\Gamma$,1). To find the entanglement wedge for this region we would have to construct a separate extremal surface that is confined to one side of the horizon and has endpoints at $\Gamma$ and $1$.} The punchline is that Cixin can use correlations between the right and left sides of the geometry to reconstruct a larger region of the \emph{right} side of the dual bulk geometry than Bindu, but only up until the Page time.

Neither group can reconstruct the entanglement wedge for $\mathcal{R}$ on the boundary. Cixin can always reconstruct the entanglement wedge of the complement $\bar{\mathcal{R}}$ that we will call the \textit{entanglement co-wedge} of $\mathcal{R}$. This means she can never reconstruct the island region on the brane. Cixin can reconstruct the more encompassing co-wedge bounded by the HM surfaces. Since Bindu has access to only one side, her only option is to reconstruct the co-wedge bounded by the island surfaces on the right side of the bulk. After the Page time, which will be zero when $\Gamma$ lies inside the constant entropy belt, the island surfaces are dominant and the island region cannot be reconstructed. However Cixin and Bindu can regain access to the island region by increasing their value of $\Gamma$, and both have access to the same information in the limit where $\Gamma\rightarrow1$.

There are some special considerations for higher-dimensional ($d > 2$) finite temperature models. The most immediate implications concern the critical angle. For the black string, when the brane is below the critical angle, the island $\mathcal{I}$ cannot extend beyond the critical anchor on the brane. In other words, the islands are confined to the atoll. This means the complementary region on the brane, $\bar{\mathcal{I}}$, can always be reconstructed when the brane lies below the critical angle. This information must be ``stored" on the defect itself because the co-wedge will always contain $\bar{\mathcal{I}}$ in the limit where $\Gamma \rightarrow 0$. This is not true above the critical angle, since the co-wedge then vanishes in the limit where $\Gamma \rightarrow 0$.\footnote{In other words, using the language in our previous paper \cite{Geng:2020fxl}, we have a tiny island surface in the limit as $\Gamma\rightarrow0$ above the critical angle.} We suspect similar considerations will hold in other finite temperature models.

It is interesting to consider the limit as $\Gamma\rightarrow1$, where the area difference diverges to infinity according to the Russian doll rule. We need $\Gamma\rightarrow1$ to reconstruct the event horizon at late times, and the Page time goes to infinity in that limit. This could indicate that information stored near the black string's event horizon is difficult to access but also robust against the loss of correlation between the two sides induced by time evolution. Suppose for example that Cixin chooses increasingly large values of $\Gamma$ and reconstructs regions of the entanglement co-wedge which are increasingly close to the event horizon. The Russian doll rule states that these surfaces will be nested and that the area difference will increase without bound as $\Gamma\rightarrow1$. Since these larger area differences will prolong the Page time, Cixin loses access to the island regions at increasingly late times. Meanwhile the island surface coincides with the black-string horizon as $\Gamma \rightarrow 1$ as the area difference diverges to positive infinity. Therefore Cixin will need to access increasingly large regions of the boundary to reconstruct regions near the event horizon at late times, but the reconstruction can always be performed, provided $\Gamma$ is large enough.

\section{Discussion of Results \label{secdiscuss}}

We have presented our computation of subregion entanglement entropy for the doubly holographic black-string model. The system is dual to a strongly coupled BCFT living on a black-hole background at finite temperature \emph{and} to an AdS$_d$ black hole coupled to a nongravitating thermal bath. In empty AdS, which corresponds to the zero-temperature limit of this system, it is known that the bulk geometry does not support island solutions for $d > 2$ when the brane is below the critical angle. 
%for $d > 2$) when the brane is below the critical angle in When working in $d > 2$ there is no guarantee that island solutions will exist at finite temperature below the critical angle, since island solutions do not exist below the critical angle in empty space. This potentially
%So an important consistency criterion for any finite temperature model with a brane is that the presence of a black hole in the bulk should lead to the existence of islands even below the critical angle.
%islands must exist even below the critical angle.
%in all such finite temperature models featuring a black hole in the bulk.

%Indeed, in the standard island story an initially growing entanglement entropy is controlled, up until the Page time, by the extremal Hartman-Maldacena surface which travels through the Einstein-Rosen bridge connecting both sides of the thermofield double. This surface is stretched out as the Einstein-Rosen bridge expands with time, which produces a Page curve because the growth is cut off by a phase transition to an island surface ending on the brane at some finite time depending on the initial area difference. 
%The existence of the critical angle in empty space might lead to the concern that such an island surface would not exist.
%But, we have pointed out that in higher dimensions such island surfaces may not exist. 

If this feature were to persist in the presence of any finite-temperature horizon, it would pose a serious problem for unitarity. Therefore an important consistency criterion for any braneworld model is that islands must exist even below the critical angle in the presence of a finite-temperature horizon. In this study we have shown that this is precisely what happens in the black-string model: islands exist at finite temperature, albeit with limited extent bounded by a critical anchor. The islands are restricted to the atoll. These solutions are consistent with empty AdS because the critical anchor stays away from the defect below the critical angle.
%in the near boundary region because the critical anchor stays far from the defect, except for branes near the critical angle, in which case the critical anchor approaches the defect as the brane angle is increased from below. 

We have summarized the phase structure for the area differences (Figure \ref{fig:phaseDiag}) with some simple geometric rules. For a given brane angle $\theta_b$, the island surfaces are dominant on the initial time slice only when the anchoring point in the bath $\Gamma$ lies within the constant entropy belt. In other words, the initial area difference in such cases is negative. This belt starts at the defect, grows monotonically with brane angle, and extends to the horizon on the bath in the limit where $\theta_b \rightarrow \pi$. The constant entropy belt forms at almost the same brane angle where the atoll first saturates the brane.

The signed area difference, which is monotonically decreasing in $\theta_b$ (Figure \ref{fig:fixed_anchor_points}) and monotonically increasing in $\Gamma$ (Figure \ref{fig:fixed_brane_angles}), features some additional geometric structure which is captured by the Russian doll rule. Nested island surfaces can be constructed by either decreasing $\theta_b$ or increasing $\Gamma$; this pushes surfaces closer to the black-string horizon, decreases the size of the island on the brane, and increases the area difference between the island surface and the HM surface. This has interesting implications for entanglement wedge reconstruction---as described in Section \ref{subsecentwedgerecon}, an experimenter would need to access increasingly large regions of the boundary to reconstruct bulk regions increasingly near the black-string horizon at late times.

It would be interesting to determine the phase structure for other doubly holographic AdS$_d$ braneworlds with $d > 2$. There are some results which we believe will be quite general even if the geometry is changed, with the critical angle playing an important role. The atoll should include a region near the horizon on the brane, because the horizon in the bulk is an extremal surface, and above the critical angle the atoll should include a region near the defect, but the atoll need not be connected. We suspect area differences will remain monotonic in both subregion size and brane angle; reducing the brane angle increases the number of degrees of freedom on the defect, and since the defect encodes the braneworld black hole, this should increase the Page time. The reduced Page time correlates with the reduced number of degrees of freedom on the defect. We hope that such general considerations will yield insights into the dynamics of BCFTs with holographic duals and help to uncover the physics underlying the critical angle, the Page angle and other aspects of the remarkable phase structure that we have found.

\section*{Acknowledgments}
We are grateful to Christoph Uhlemann for sharing his related top-down results prior to publication. HG is very grateful to his parents and recommenders. The work of HG is supported by the grant (272268) from the Moore Foundation ``Fundamental Physics from Astronomy and Cosmology." The work of AK and MR was supported, in part, by the U.S. Department of Energy under Grant-No. DE-SC0022021 and a grant from the Simons Foundation (Grant 651440, AK). The work of CP was supported in part by the National Science Foundation under Grant No.~PHY-1914679 and by the Robert N. Little Fellowship. The work of SR was partially supported by a Swarnajayanti fellowship, DST/SJF/PSA-02/2016-17 from the Department of Science and Technology (India). Research at ICTS-TIFR is supported by the government of India through the Department of Atomic Energy grant RTI4001. The work of LR is supported by NSF grants PHY-1620806 and PHY-1915071, the Chau Foundation HS Chau postdoc award, the Kavli Foundation grant ``Kavli Dream Team,'' and the Moore Foundation Award 8342. SS was supported by National Science Foundation (NSF) Grants No. PHY-1820712 and PHY-1914679.

\begin{appendices}

%%%%%%%%%%%%%%%%%%%%%%%%%%%%%%%
\section{Analytic Results for the Area Difference in the $\Gamma \rightarrow 1$ limit}\label{app:A}
%%%%%%%%%%%%%%%%%%%%%%%%%%%%%%%

\begin{figure}
    \centering
    \includegraphics[width=.9\linewidth]{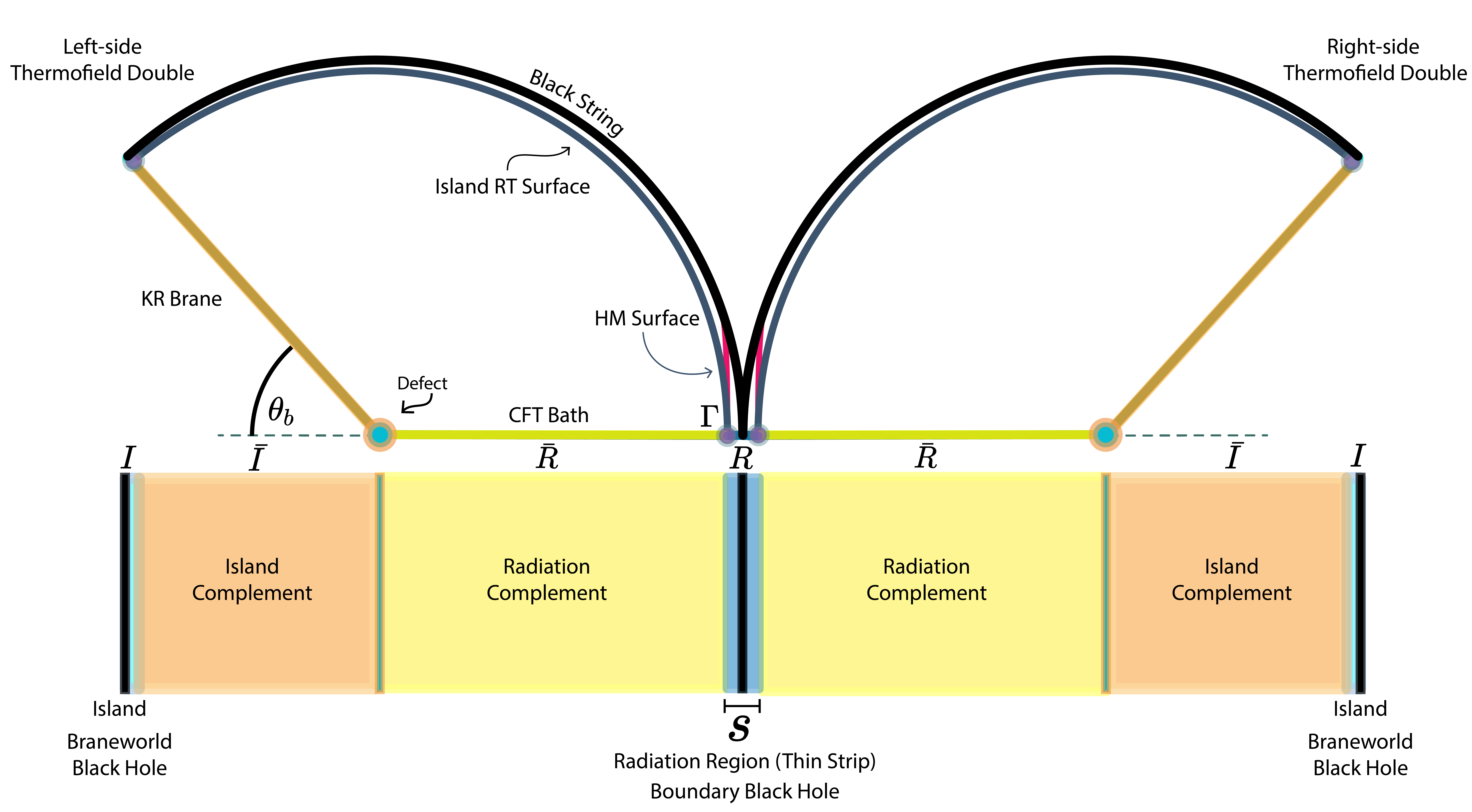}
    \caption{Here we illustrate the doubly holographic black string setup for a small radiation region of width $s$. This corresponds to the limit $\Gamma\rightarrow1$. The island region becomes infinitesimally small as well; this is consistent with the Russian doll rule.}
    \label{fig:thin_strip}
\end{figure}

In the bulk of this paper we have considered RT surfaces anchored at a given point $\Gamma$ in the nongravitating black hole bath, and an symmetric point $\tilde{\Gamma} = \Gamma$ in the second asymptotic region of the same black hole. In the limit that $\Gamma \rightarrow 1$ these two anchor points approach the bifurcation surface from both sides. Furthermore, the spacetime near the black horizon becomes essentially flat. So the initial HM surface in this limit is really calculating the entanglement entropy of a thin strip in empty AdS, illustrated in Figure \ref{fig:thin_strip}, for which analytic results have been known for a long time \cite{Ryu:2006ef}.

The entanglement entropy of a strip of length $s$ in a nongravitating $\text{CFT}_{d}$ with dual empty $\text{AdS}_{d+1}$ is given by \cite{Ryu:2006ef},
\begin{equation}
 S = \frac{1}{4G_{N}^{(d+1)}} \left[ \frac{2}{d-2} \left( \frac{L}{\epsilon} \right)^{d-2} - \frac{2^{d-1} \pi^{\frac{d-1}{2}}}{d-2} \left( \frac{\Gamma\left(\frac{d}{2(d-1)}\right)}{\Gamma\left(\frac{1}{2(d-1)} \right)} \right)^{d-1} \left( \frac{L}{s} \right)^{d-2} \right], \label{eq:TakaS}
\end{equation}
where the first term is the divergent piece, which is subtracted upon renormalization. $L^{d-2}$ is the volume factor of to the orthogonal directions which were generally omitted in this note. Since the near-horizon geometry is flat, in order to compare with the above expression we need to transform our coordinate $u=u_{h}-\zeta$, so that,
\begin{equation}
 \frac{du^{2}}{h(u)} = \frac{du^{2}}{1-\frac{u^{d-1}}{u_{h}^{d-1}}} \approx \frac{d\zeta^{2}}{\frac{d-1}{u_h} \zeta}.
\end{equation}
Upon transforming the coordinate, we take the near-horizon limit. In these new coordinates, we define $\zeta = \zeta_*$ to correspond to $u = u_h \Gamma$.

Now we define the flat coordinate such that $dy = d \zeta/\sqrt{\zeta\frac{d-1}{u_h}}$. The length of the near-horizon belt in the flat setup is $s = 2 \int_0^{y_*} dy = 2 y_*$, where the factor of 2 comes from the fact that our geometry is replicated on the other side of the horizon. In this coordinate, we have  $y_* = 2\sqrt{\zeta_* \frac{u_h}{d-1}}$. Now, substituting this into $\eqref{eq:TakaS}$, dividing by the transverse volume factor $L^{d-2}$, and renormalizing the expression we obtain,
\begin{equation}
2\mathcal{A}_{HM} = -\frac{2^{d-1} \pi^{\frac{d-1}{2}}}{d-2} \left( \frac{\Gamma\left(\frac{d}{2(d-1)}\right)}{\Gamma\left(\frac{1}{2(d-1)} \right)} \right)^{d-1} \left( \frac{1}{4} \sqrt{\frac{d-1}{\zeta_* u_{h}}} \right)^{d-2},
\end{equation}
where we have multiplied the entropy by $4 G_N^{(d+1)}$ to obtain the corresponding area. However, note that this is the area of the \textit{full} initial HM surface spanning both exterior regions of the bulk black string; to appropriately match with the conventions of this paper, we use $\mathcal{A}$ to mean the area of the HM surface restricted to a single exterior region of the bulk.

\begin{figure}
 \centering
 \includegraphics[scale=0.35]{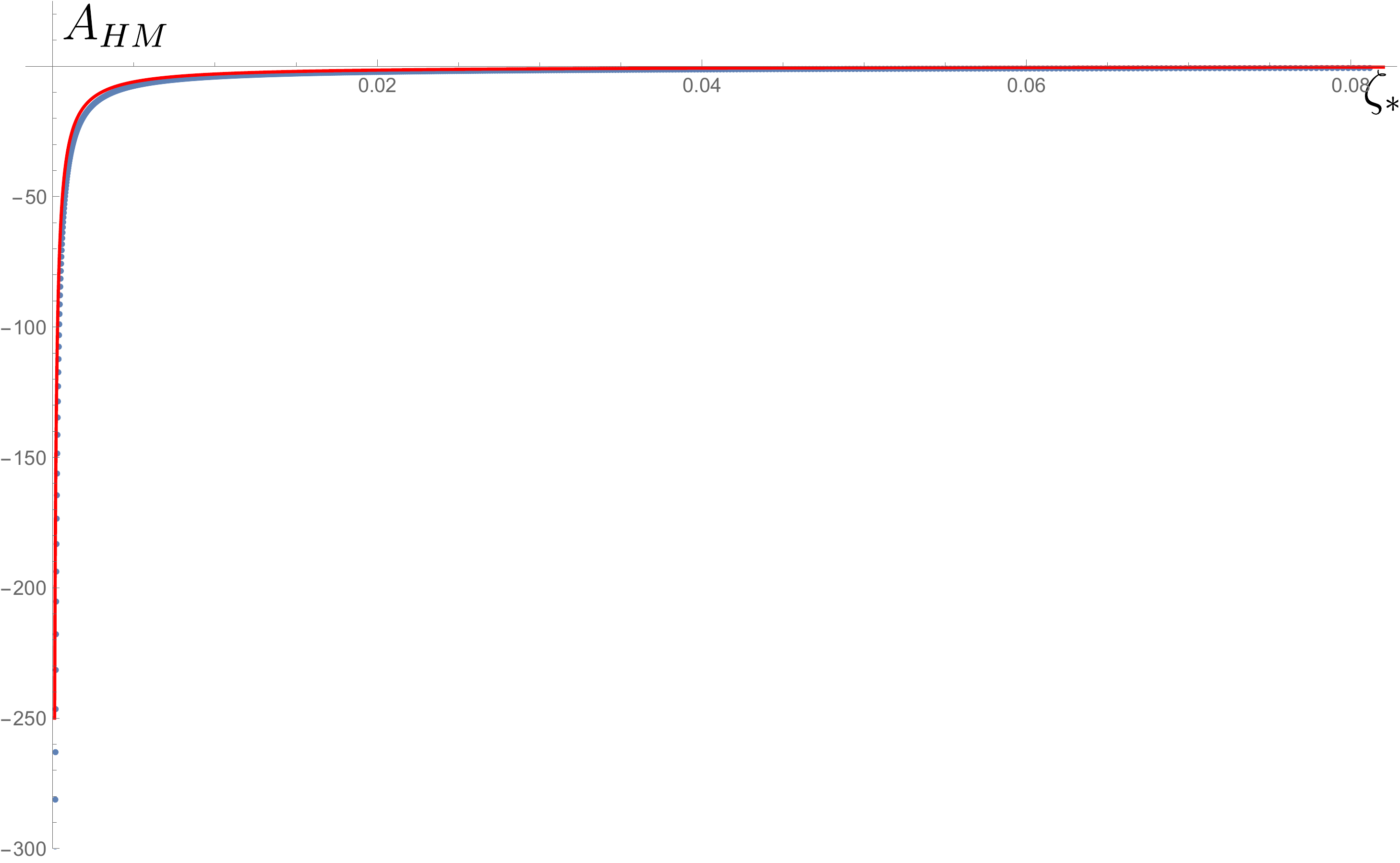}
 \caption{Area of the HM surface as a function of $\zeta_*$ with $u_h = 1$. The red curve is what we expect for the near-horizon region (small $\zeta_*$) from theory, namely $-0.030/\zeta_*$. Close to the horizon, the data roughly matches the theory, but for some large enough $\zeta_*$ the finite temperature kicks in and we should not expect the data to match. It is not visible, but there is a small error in the ``large" $\zeta_*$ region.}
 \label{fig:AHMnear1}
\end{figure}

In the limit where $\Gamma \rightarrow 1$, we take the above expression to $\zeta_* \rightarrow 0$. After substituting $d=4$ and $u_h=1$, we find that,
\begin{equation}
\label{analyticform}
\mathcal{A}_{HM} = -\frac{3\pi^{3/2}\Gamma\left(\frac{2}{3}\right)^3}{8\Gamma\left(\frac{1}{6}\right)^3} \frac{1}{\zeta_*} \approx - \frac{0.030}{\zeta_*}.
\end{equation}
We plot our numerical results for the Hartman-Maldacena area close to $\zeta_* = 0$ in Figure \ref{fig:AHMnear1} and demonstrate that they agree well with the analytic form \eqref{analyticform}, validating our method.

%%%%%%%%%%%%%%%%%%%%%%%%%%%%%%%%%%%%%%%%%%%%%%%%%%%%%%
\section{Validation of Numerics} \label{app:B}
%%%%%%%%%%%%%%%%%%%%%%%%%%%%%%%%%%%%%%%%%%%%%%%%%%%%%%

\begin{figure}
 \centering
 \includegraphics[scale=0.3875]{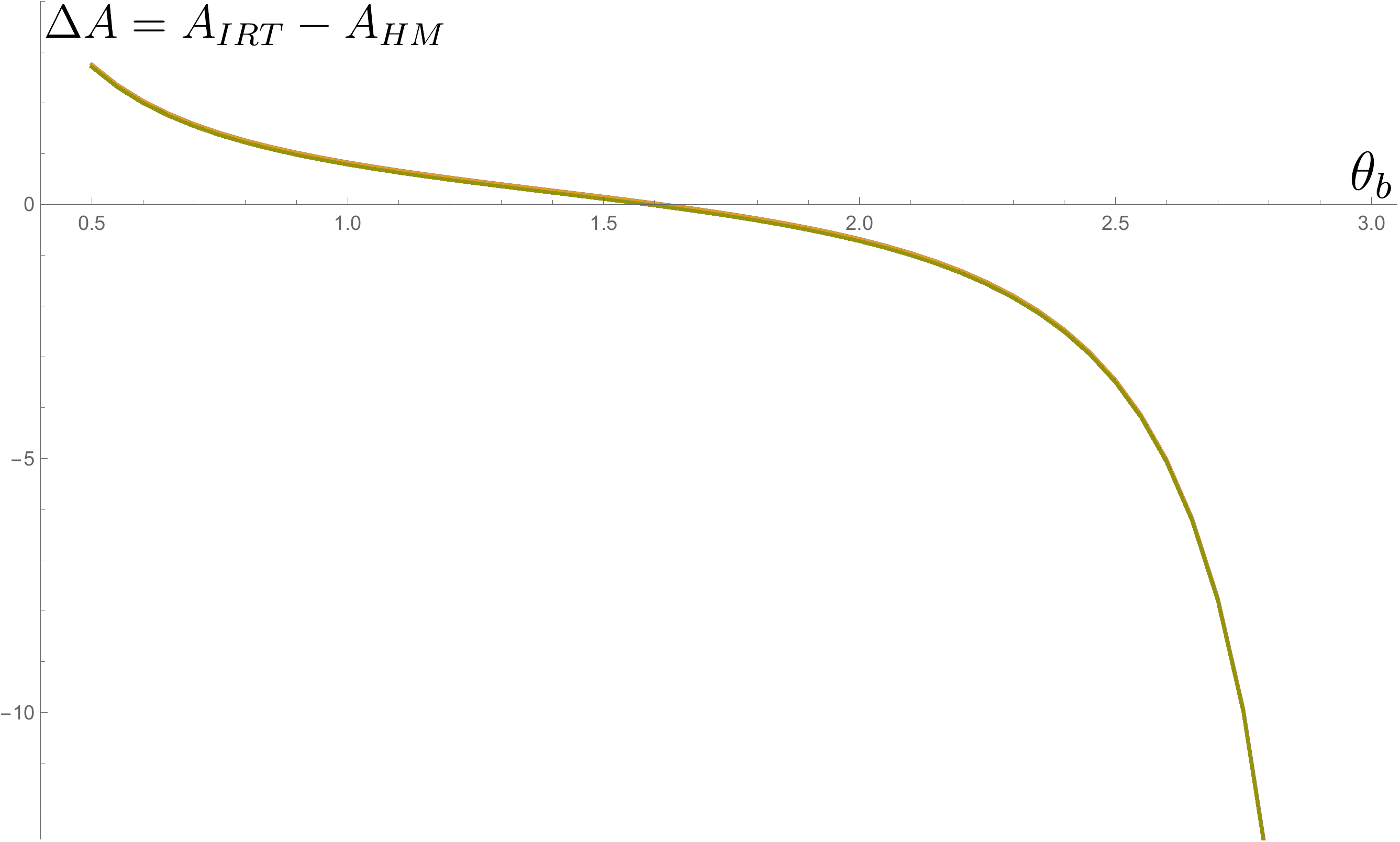}
 \caption{Area difference as a function of brane angle, for $\Gamma = 0.32$ for angular cutoffs ranging from 0.0005 to 0.2 in steps of 0.0005 (with $u_h = 1$). Each curve is shown in a different colour, but since they are very close it is hard to distinguish one from the other.}
 \label{fig:CutoffDep}
\end{figure}

One may wonder whether the results we are obtaining are indeed physical. To show that this is the case, we compute the area difference as a function of the brane angle for different cutoffs. We see in Figure $\ref{fig:CutoffDep}$ that our area differences do not vary substantially for a large range of cutoff choices.

\end{appendices}

\bibliographystyle{JHEP}
\bibliography{references}
\end{document}